\documentclass[a4paper,fleqn,usenatbib,useAMS]{mnras}

\usepackage{graphicx}	
\usepackage{amsmath}	
\usepackage{amssymb}	
\usepackage{multicol}        
\usepackage{bm}		
\usepackage{pdflscape}	
\usepackage{here}

\usepackage[T1]{fontenc}
\usepackage{ae,aecompl}

\usepackage{newtxtext,newtxmath}

\title[Dust Simulation with size distribution]{Galaxy simulation with the evolution of grain size distribution}

\author[S. Aoyama et al. ]{Shohei Aoyama,$^{1}$\thanks{Contact e-mail: \href{mailto:aoyama.naoj@gmail.com}{aoyama.naoj@gmail.com}}
Hiroyuki Hirashita,$^{1}$ and Kentaro Nagamine$^{2,3,4}$\\
$^{1}$ Institute of Astronomy and Astrophysics, Academia Sinica,
Astronomy-Mathematics Building, AS/NTU,
No.\ 1, Sec.\ 4, Roosevelt Road, Taipei 10617, Taiwan\\
$^{2}$ Theoretical Astrophysics, Department of Earth \& Space Science, Osaka University, 1-1 Machikaneyama, Toyonaka, Osaka 560-0043, Japan\\
$^{3}$ Kavli IPMU (WPI), The University of Tokyo, 5-1-5 Kashiwanoha, Kashiwa, Chiba, 277-8583, Japan \\
$^{4}$ Department of Physics \& Astronomy, University of Nevada, Las Vegas, 4505 S. Maryland Pkwy, Las Vegas, NV 89154-4002, USA}

\date{Last updated ***}

\pubyear{2019}

\begin{document}
\label{firstpage}
\pagerange{\pageref{firstpage}--\pageref{lastpage}}
\maketitle
\begin{abstract}
We compute the evolution of interstellar dust in a hydrodynamic simulation 
of an isolated disc galaxy.
We newly implement the evolution of full grain size distribution by
sampling 32 grid points on the axis of
the grain radius. We solve it consistently with the chemical enrichment and hydrodynamic evolution of the galaxy.
This enables us to theoretically investigate spatially resolved evolution of grain size distribution in a galaxy.
The grain size distribution evolves from a large-grain-dominated
($\gtrsim 0.1~\micron$) phase to a small-grain production phase, eventually converging to
a power-law-like grain size distribution similar to the so-called MRN distribution. 
We find that the small-grain abundance is higher in the dense ISM in the early epoch ($t\lesssim 1$~Gyr) because of efficient dust growth by accretion, while coagulation makes the small-grain abundance less enhanced in the dense ISM later. This leads to steeper extinction curves in the dense ISM than in the diffuse ISM in the early phase, while they show the opposite trend later.
The radial trend
{of extinction curves}
is described by faster evolution in the inner part.  We also confirm that the simulation reproduces the observed relation between dust-to-gas ratio and metallicity, and the radial gradients of dust-to-gas ratio and dust-to-metal ratio 
{in nearby galaxies}. 
Since the above change in the grain size distribution occurs in $t\sim 1$ Gyr, the age and density dependence of grain size distribution has a significant impact on the extinction curves even at high redshift.
\end{abstract}





\section{Introduction}
Dust grains contained in the interstellar medium (ISM) evolve
together with galaxies. Dust plays an important role at
various stages of galaxy evolution in the following aspects.
Molecular hydrogen, which is the main constituent of molecular clouds, forms
on dust surfaces. Thus, dust enrichment induces an environment
rich in molecular clouds that eventually host star formation
\citep[][]{2002MNRAS.337..921H,2004ApJ...604..222C,2011ApJ...735...44Y}.
In addition, dust also acts as a coolant in the late stages of star
formation, inducing the fragmentation of
gas clouds \citep{2005ApJ...626..627O}
{. This}
determines the typical mass of stars \citep[][]{2006MNRAS.369.1437S} and the shape
stellar initial mass function (IMF) \citep[e.g.][]{2012MNRAS.419.1642C}.
All the above processes occur on grain surfaces.
The total grain surface area is governed not only by the grain abundance
but also by the distribution function of grain size (grain size distribution);
therefore, the evolution of these two aspects
of dust grains is fundamentally important for the thorough understanding
of galaxy evolution.

Dust is also important in radiative processes.
Dust grains absorb ultraviolet (UV)--optical stellar light
and reemit infrared (IR) photons. Thus,
the spectral energy distribution (SED) is dramatically changed
by dust grains \citep[e.g.][]{2005A&A...440L..17T}.
This means that a part of star formation activities in galaxies
are obscured by dust. This obscured fraction is only recovered by
the observations of IR dust emission; thus, tracing the dust emission is
crucial in estimating the total star formation rate (SFR)
\citep[e.g.][]{1996A&A...306...61B, 2000PASJ...52..539I,2012ARA&A..50..531K}.
Because the absorption efficiency of UV photons depends strongly 
on the grain size distribution through the extinction curve
as well as on the total dust abundance \citep[e.g.][]{2001PASP..113.1449C},
the two aspects of dust
(namely, the total dust abundance and the grain size distribution) are
important in understanding galaxy evolution.

A comprehensive model that computes the evolution of grain size distribution
has been developed by \cite{2013MNRAS.432..637A}, 
who took into account all major processes
that drive the change of grain size distribution in a one-zone model of a galaxy.
Dust condenses from the metals ejected from supernovae (SNe)
\citep[e.g.][]{1989ApJ...344..325K,2001MNRAS.325..726T,2003ApJ...598..785N,2007MNRAS.378..973B,2007ApJ...666..955N}
and asymptotic giant branch (AGB) stars
\citep[e.g.][]{2006A&A...447..553F, 2014MNRAS.439..977V, 2017MNRAS.467.4431D}.
This dust production by stellar sources is the starting point of dust evolution.
Calculations of dust condensation and processing in stellar ejecta indicate that
grains are biased to large sizes ($a\gtrsim 0.1~\micron$, where $a$ is the grain radius)
\citep[][]{2007ApJ...666..955N, 2012ApJ...745..159Y, 2017MNRAS.467.4431D}.
Subsequently, dust grains are processed in the ISM as described by
\cite{2013MNRAS.432..637A} and \citet[][hereafter HA19]{2019MNRAS.482.2555H}.
The interstellar processing
starts with shattering, which gradually converts large grains into small ones. 
As the system is enriched with metals, the accretion of gas-phase metals onto
dust grains becomes efficient in the dense ISM. This causes a drastic
increase of { the grain abundance at $a\lesssim 0.03~\micron$}
because small grains have
larger surface-to-volume ratios (recall that the accretion rate is proportional to the grain
surface area). Accretion is saturated when a significant fraction of gas-phase metals
are used up; and subsequently, coagulation converts small grains into large grains,
smoothing the grain size distribution towards a power-law shape similar to the
\citet[][hereafter MRN]{1977ApJ...217..425M} grain size distribution.

In reality, the evolution of dust is strongly affected by the ambient physical condition
such as gas metallicity, temperature, and density.
Although \cite{2013MNRAS.432..637A} succeeded in calculating the evolution 
of grain size distribution, their one-zone model is not capable of
treating the spatial variation of dust evolution that depends on the ISM conditions. 
The evolution of ISM is governed by hydrodynamical processes.
Thus, hydrodynamic simulations have been used to study the evolution 
of ISM in galaxies
\citep[e.g.][]{2007ApJ...660..276W,2008PASJ...60..667S,2014MNRAS.442.3745M,2014ApJS..210...14K,2016ApJ...833..202K,2019arXiv190508806M, 2019MNRAS.490.1425L, 2019arXiv190907388G}.
For the purpose of considering the effects of ISM evolution on dust properties,
it would be ideal to treat dust evolution in a manner consistent with
the hydrodynamical development of the ISM. However,
it is in general computationally expensive to implement the evolution
of grain size distribution in already heavy hydrodynamic simulations.

In the past, most hydrodynamic simulations with dust evolution
calculated only the total dust abundance without grain size distribution.
\cite{2010MNRAS.403..620D} post-processed
a cosmological hydrodynamic simulation with a dust evolution model
which includes stellar dust production and SN destruction, and 
calculated the expected submillimetre fluxes for high-redshift star-forming galaxies.
\cite{2015MNRAS.451..418Y}, using a cosmological zoom-in simulation, 
calculated the dust distribution in high-redshift galaxies by simply 
assuming a constant dust-to-metal ratio.
\cite{2015MNRAS.449.1625B} treated dust as a new particle species 
in addition to the gas, dark matter and star particles 
in their hydrodynamic simulations. In order to
calculate the evolution of dust in galaxies,
they included not only the formation and destruction of dust, but also
the effect of dust on star formation and stellar feedback.
\cite{2016MNRAS.457.3775M} performed cosmological zoom-in	
simulations by modelling dust as a component coupled with the gas.
They included the processes relevant for the formation and evolution of dust,
and revealed the importance of dust growth by the accretion of gas-phase metals. 
In addition, \cite{2017MNRAS.468.1505M} ran cosmological simulations 
to calculate the statistical properties of dust in galaxies. 
They broadly reproduced the dust abundances in the present-day galaxies, 
although they tended to underestimate those in high-redshift dusty galaxies.
\cite{2016ApJ...831..147Z} analysed the dust evolution in 
an isolated Milky-Way-like galaxy by post-processing a hydrodynamic simulation. 
They put particular focus on the effect of
gas-temperature-dependent sticking coefficient on dust growth by accretion,
in order to reproduce the relation between silicon depletion and gas density.
\cite{2019MNRAS.487.3252H}
simulated the dust evolution in the ISM, focusing on dust destruction
by sputtering. Although they did not trace the entire galaxy, their focus on
a small region in the ISM enabled them to resolve the structures associated with SN shocks.
We also note that there are some non-hydrodynamic approaches 
using semi-analytic models that predicted the evolution of dust mass in galaxies
\citep[][]{2017MNRAS.471.3152P, 2014MNRAS.445.3039D, 2018MNRAS.473.4538G,2019MNRAS.489.4072V}.

\citet[][hereafter A17]{2017MNRAS.466..105A} implemented a two-size grain model
in the cosmological smoothed particle hydrodynamics (SPH) code \textsc{gadget3-osaka}
(A17; \citealt{2019MNRAS.484.2632S}),
which is a modified version 
of \textsc{gadget-3}
\citep[originally described in][as \textsc{gadget-2}]{2005MNRAS.364.1105S}. 
Their treatment of dust is based on the \textit{two-size approximation}
\citep{2015MNRAS.447.2937H}, 
in which the entire grain radius range is divided into `large' and `small' grains.
The two-size approximation reduces the computational cost in calculating the
evolution of grain size distribution,
so that it is easily implemented in hydrodynamic simulations. 
A17 performed simulations of a Milky-Way-like isolated galaxy, and 
explained the radial distribution of the total dust abundance
of nearby galaxies in \cite{2012MNRAS.423...38M}.
A17 also predicted a variation of grain size distribution along the galactic radius. 
\cite{2017MNRAS.469..870H} took advantage of the grain size information in A17 
and investigated the temporal and spatial variation of extinction curves.
They also examined the dependence of extinction curves on metallicity.
They found that extinction curves become steeper at
the intermediate age and metallicity when the dust abundance is strongly
modified by the accretion of gas-phase metals, and at low densities 
where shattering is more efficient than coagulation.
Using the same simulation, \cite{2018MNRAS.474.1545C} calculated the
abundance of H$_2$ and CO in a consistent manner with the dust abundance and
grain size distribution, which govern the grain-surface reaction rates and UV shielding.
They found that H$_2$ is not a good tracer of SFR at low metallicity
because H$_2$-rich regions are limited to dense compact regions.

The above framework with the two-size approximation 
has also been applied to cosmological simulations
\citep[][]{2018MNRAS.478.4905A, 2018MNRAS.479.2588G, 2019MNRAS.484.1852A, 2019MNRAS.485.1727H}.
For example, \cite{2018MNRAS.478.4905A}
explained the observed relation between dust-to-gas ratio and metallicity,
the evolution of the comoving dust mass density in the Universe, and
the radial profile of dust surface density around massive galaxies.
\cite{2019MNRAS.485.1727H}, using the same simulation setup but including
a simple treatment for active galactic nucleus feedback, 
focused on the relations between dust-related quantities and galaxy characteristics. 
They broadly reproduced the statistical properties such as the dust mass function
and the relation between dust-to-gas ratio and metallicity in the nearby Universe. 
They also investigated extinction curves.
However, because these results are based on the two-size approximation, 
the accuracy of extinction curve is limited.
{There is another approach for the evolution of
  grain size distribution based on the method of moments formulated by \cite{MATTSSON2016107}. This method is powerful in calculating the processes related to the moments of
grain size distribution such as the 
total grain surface area.
However, 
  it is still difficult to calculate the extinction curves.
  An explicit solution for the full grain size distribution enables us to
  directly calculate the extinction curves as well as other quantities
  (e.g.\ the surface area) that require grain size information.
}

Recently, there have been some efforts to directly solve the evolution of full
grain size distribution in galaxy-scale hydrodynamic simulations.
\cite{2018MNRAS.478.2851M} developed a computational code that solves
the evolution of full grain size distribution in an adaptive mesh hydrodynamic
simulation code, {\tt AREPO} \citep[][]{2010MNRAS.401..791S}.
They showed some test calculations of an isolated galaxy, 
but realistic calculations to be compared with
observations need further implementation of, for example, stellar feedback.
\citet[][hereafter HA19]{2019MNRAS.482.2555H} took another approach to solve the evolution of full
grain size distribution. They post-processed an isolated-galaxy simulation:
they chose some SPH gas particles
and derived the evolution of grain size distribution on each particle based on
its history of physical conditions (mainly gas density, temperature, and metallicity).
They successfully derived the grain size distribution similar to
the MRN distribution at ages comparable to the Milky Way ($\gtrsim 3$~Gyr).
However, because their calculation was based on post-processing,
the results may depend on the frequency of snapshots; that is,
we cannot capture the change of physical conditions shorter than the
output snapshot interval ($10^{7}$~yr in the above post-processing).
Indeed, the processes occurring in dense environments such as
coagulation and accretion could have a shorter characteristic time-scales
in metal-rich regions (A17).
To obtain robust results against such
potentially quick processes, we need to solve the evolution of grain
size distribution and that of hydrodynamical structures simultaneously.

The goal of this paper is to implement the evolution of full grain size distribution
directly in a hydrodynamic simulation to obtain a self-consistent view of
grain size distribution with ISM evolution. We use an isolated-galaxy simulation,
which could achieve a higher spatial resolution of the ISM compared with
cosmological simulations. Two improvements
with respect to previous studies are expected: one is direct knowledge of
full grain size distribution without relying on the two-size approximation. This leads to
a better understanding of the quantities for which the grain size distribution is
essential (such as extinction curves). The other is a better capability of treating
the dust growth processes (accretion and coagulation), which could occur
quickly. Such quick processes cannot be captured robustly by post-processing so that a
simultaneous calculation of grain size distribution and hydrodynamics is essential.

This paper is organized as follows. 
In Section \ref{sec:model}, we introduce the calculation method for the
grain size distribution and the scheme of the simulation.  
In Section \ref{sec:result}, we show the results and comparison with observations.
In Section \ref{sec:discussion}, we provide additional discussions.
Section \ref{sec:conclusion} concludes this paper.
Throughout this paper, we adopt the solar metallicity
$Z_{\sun}=0.02$ for the convenience
of direct comparison with our previous simulations. 
We also use the terms `small' and `large' grain radius to roughly
indicate the grain radius smaller and larger than $\sim 0.03~\micron$.


\section{Model}\label{sec:model}

\subsection{Hydrodynamic simulation}

We perform a hydrodynamic simulation of an isolated galaxy
using the smoothed particle hydrodynamics (SPH) code 
\textsc{gadget3-osaka} \citep[A17;][]{2019MNRAS.484.2632S}, 
which is based on the \textsc{gadget-3} code \citep[originally described in][]{2005MNRAS.364.1105S}. 
In our simulation, the star formation and metal/dust production are treated consistently
with gravitational/hydrodynamic evolution of a Milky-Way-like isolated galaxy.
Dust grains are assumed to be coupled with gas particles (Section \ref{subsec:size_distribution}).

Our present run corresponds to the M12 run in \citet{2019MNRAS.484.2632S},  and the total mass of isolated galaxy is $\sim 10^{12}~{\rm M}_{\sun}$.
We employ $10^5$ dark matter particles, $10^5$ gas (SPH) particles, $10^5$ and $1.25\times 10^4$ collisionless particles that represent stars in the disc and the bulge, respectively.  We adopt a fixed gravitational softening length of $\epsilon_{\rm grav} = 80$\,pc, but allow the minimum gas smoothing length to reach 10 percent of $\epsilon_{\rm grav}$.  The final gas smoothing length reaches $\sim 30$\,pc as the gas becomes denser owing to radiative cooling.

The difference between this paper and A17 is as follows. 
We additionally include the metal/dust supply and the feedback from AGB stars. 
We use the {\small CELib} package \citep[][]{2017AJ....153...85S}
for metal generation and follow 13 chemical abundances 
(H, He, C, N, O, Ne, Mg, Si, S, Ca, Fe, Ni, and Eu). 
The ejection of metals and energy from stars are treated separately 
for SNe Ia, II, and AGB stars as a function of time after the onset of star formation.  
The delay-time distribution function of SN Ia rate is modeled with a power-law of $t^{-1}$
\citep[][]{2008PASJ...60.1327T, 2012PASA...29..447M}.
In comparison, in A17, 
the metal and dust were injected instantaneously after 4\,Myr from the onset of star formation, whose masses were simply proportional to the stellar mass.
For more details of the stellar feedback with CELib in our simulations, 
we refer the interested reader to \citet{2019MNRAS.484.2632S}.
We further describe the dust enrichment by stellar sources in the next subsection.

\subsection{Evolution of grain size distribution}\label{subsec:size_distribution}

We use the same model as in HA19 for the evolution of grain size distribution.
The model is based on \cite{2013MNRAS.432..637A}, but with simplifications
that do not lose the physical essence
of the relevant processes. These simplifications are useful for the implementation in
hydrodynamic simulations.
For the dust evolution processes,
we consider stellar dust production, dust destruction by SN shocks in
the ISM, dust growth by accretion and coagulation in the dense ISM, and
dust disruption by shattering in the diffuse ISM. We assume that the dust grains
are dynamically coupled with the gas, which is usually valid on the spatial scales
of interest in this paper \citep[][]{2018MNRAS.478.2851M}.
In what follows, we only describe the outline of the model to be implemented for
individual SPH gas particles (hereafter, referred to simply as gas particles).
We refer the interested reader to HA19 for the complete set of equations and
the details.

The grain size distribution at time $t$ is expressed by the grain mass distribution
$\rho_\mathrm{d}(m,\, t)$, which is defined such that
$\rho_\mathrm{d}(m,\, t)\,\mathrm{d}m$
($m$ is the grain mass and $t$ is the time) is the mass density
of dust grains whose mass is between $m$ and $m+\mathrm{d}m$.
We assume grains to be spherical and compact,
so that $m=(4\upi /3)a^3s$, where $a$ is the grain radius and $s$
is the material density of dust.
We adopt $s=3.5$ g cm$^{-3}$ based on silicate in
\cite{2001ApJ...548..296W}.
The grain mass distribution is related
to the grain size distribution,
$n(a,\, t)$, as
\begin{align}
\rho_\mathrm{d}(m,\, t)\,\mathrm{d}m=\frac{4}{3}\upi a^3sn(a)\,\mathrm{d}a.
\label{eq:rho_n}
\end{align}
The total dust mass density $\rho_\mathrm{d,tot}(t)$ is
\begin{align}
\rho_\mathrm{d,tot}(t)=\int_0^\infty\rho_\mathrm{d}(m,\, t)\,\mathrm{d}m.
\end{align}
The gas density, $\rho_{\rm gas}$, is given by the number density of hydrogen nuclei,
$n_\mathrm{H}$ as
$\rho_\mathrm{gas}=\mu m_\mathrm{H}n_\mathrm{H}$
($\mu =1.4$ is the gas mass per hydrogen, and $m_\mathrm{H}$ is the mass
of hydrogen atom). The dust-to-gas ratio, $\mathcal{D}(t)$, is defined as
\begin{align}
\mathcal{D}_\mathrm{tot}(t)\equiv\frac{\rho_\mathrm{d,tot}(t)}{\rho_\mathrm{gas}(t)}.\label{eq:dg}
\end{align}
Note that the evolution of $\rho_\mathrm{gas}$ is calculated by
the hydrodynamic simulation for each gas particle.

The time-evolution of $\rho_\mathrm{d}(m,\, t)$ is described by
\begin{align}
\frac{\upartial\rho_\mathrm{d}(m,\, t)}{\upartial t} &=
\left[\frac{\upartial\rho_\mathrm{d}(m,\, t)}{\upartial t}\right]_\mathrm{star}+
\left[\frac{\upartial\rho_\mathrm{d}(m,\, t)}{\upartial t}\right]_\mathrm{sput}
\nonumber\\
&+
\left[\frac{\upartial\rho_\mathrm{d}(m,\, t)}{\upartial t}\right]_\mathrm{acc}+
\left[\frac{\upartial\rho_\mathrm{d}(m,\, t)}{\upartial t}\right]_\mathrm{shat}
\nonumber\\
&+
\left[\frac{\upartial\rho_\mathrm{d}(m,\, t)}{\upartial t}\right]_\mathrm{coag}
+\rho_\mathrm{d}(m,\, t)\frac{\mathrm{d}\ln\rho_\mathrm{gas}}{\mathrm{d}t},
\label{eq:basic}
\end{align}
where the terms with subscripts `star', `sput', `acc', `shat' and `coag'
indicate the changing rates of grain mass distribution by stellar dust
production, sputtering, accretion, shattering, and coagulation,
respectively, and the last term
is caused by the change of background gas density.
Below we briefly describe each term.
We actually solve discrete forms and adopt 32 grid points 
for the discrete grain size distribution in the range of
$3\times 10^{-4} -10\,\micron$.

\subsubsection{Stellar dust production}\label{subsec:stellar}

We estimate the increase of dust mass by stellar dust production, assuming
a constant dust condensation efficiency, $f_\mathrm{in}=0.1$, of the
ejected metals
\citep[][]{2011EP&S...63.1027I,2013MNRAS.436.1238K}.
{Using the metal production rate per volume, $\dot{\rho}_{\rm Z}$, calculated in the simulations,}
we write the change of the grain size distribution by stellar dust production as
\begin{align}
\left[\frac{\upartial\rho_\mathrm{d}(m,\, t)}{\upartial t}\right]_\mathrm{star}=
f_\mathrm{in}\dot{\rho}_Z\, m\tilde{\varphi} (m),\label{eq:stellar}
\end{align}
where $m\tilde{\varphi} (m)$ is the mass distribution function of
the dust grains produced by stars, and it is normalized so that the integration
for the whole grain mass range is unity. The grain size distribution of
the dust grains produced by stars [which is related to the above mass
distribution as $\varphi (a)\,\mathrm{d}a\equiv\tilde{\varphi}(m)\,\mathrm{d}m$]
is written as
\begin{align}
{\varphi}(a)=\frac{C_\varphi}{a}\exp\left\{
-\frac{[\ln (a/a_0)]^2}{2\sigma^2}\right\} ,
\end{align}
where $C_\varphi$ is the normalization factor, $\sigma$ is the standard deviation,
and $a_0$ is the central
grain radius. We adopt $\sigma =0.47$ and $a_0=0.1\,\micron$ following
\cite{2013MNRAS.432..637A}.

\subsubsection{Dust destruction and growth}\label{subsubsec:destruction}

As shown by HA19, dust destruction by sputtering in SN shocks and dust growth
by accretion are both described by the following same form of equation: 
\begin{align}
\left[\frac{\upartial\rho_\mathrm{d}(m,\, t)}{\upartial t}\right]_\mathrm{sput/acc}
=-\frac{\upartial}{\upartial m}[\dot{m}\rho_\mathrm{d}(m,\, t)]+
\frac{\dot{m}}{m}\rho_\mathrm{d}(m,\, t).\label{eq:sputtering}
\end{align}
where $\dot{m}\equiv\mathrm{d}m/\mathrm{d}t$. 
We estimate that
\begin{align}
\dot{m}=\xi (t)m/\tau_\mathrm{dest/acc}(m),
\end{align}
where $\tau_\mathrm{dest/acc}(m)$ is the grain-mass-dependent time-scale of
destruction or accretion.
In the actual computations, we solve the effects of destruction and accretion,
separately.
For destruction, we formally choose $\xi (t)=-1$.
For accretion, $\xi (t)$ is the fraction of metals in the gas-phase
[i.e.\ the fraction $(1-\xi )$ is condensed into the dust phase].
Therefore,
\begin{align}
  \xi (t) \equiv
  \begin{cases}
    1-\mathcal{D}_{\rm tot}\slash Z\,\,&{(\rm accretion)},\\
    -1\,\, &{(\rm destruction)},
\end{cases}
\end{align}
where $Z$ is the metallicity.

The destruction time-scale $\tau_\mathrm{dest}(m)$ is estimated based on
the sweeping time-scale divided by the grain-mass-dependent destruction
efficiency as
\citep[e.g.][]{1989IAUS..135..431M}
\begin{align}
\tau_\mathrm{dest}(m)=
\frac{M_\mathrm{gas}}{\epsilon_\mathrm{dest}(m)M_\mathrm{s}\gamma},
\label{eq:tau_dest}
\end{align}
where $M_\mathrm{gas}$ is the mass of the gas particle of interest,
$M_\mathrm{s}=6800$~M$_{\sun}$ is the gas mass swept by a single SN
blast, $\gamma$ is the rate of SNe sweeping the gas,
$\epsilon_{\rm dest} (m)$ is the dust destruction efficiency as a function of grain mass.
We adopt the following empirical expression for the destruction efficiency
(described as a function of $a$ instead of $m$):\footnote{
{We corrected a typo in HA19 for this equation.
}
  }
\begin{align}
\epsilon_{\rm dest}(a)=1-\exp\left[-0.1\left(\frac{a}{0.1\,\micron}\right)^{-1}\right] .
\label{eq:eps}
\end{align}

For accretion, the growth time-scale is
estimated as (we fixed the sticking efficiency $S=0.3$ in HA19)
\begin{align}
\tau_\mathrm{acc}(m) &= \frac{1}{3} \tau^\prime_\mathrm{0,acc}\left(\frac{a}{0.1~\micron}
\right)\left(\frac{Z}{\mathrm{Z}_{\sun}}\right)^{-1}
\left(\frac{n_\mathrm{H}}{10^3\,\mathrm{cm}^{-3}}
\right)^{-1}\left(\frac{T_\mathrm{gas}}{10\,\mathrm{K}}
\right)^{-1/2},
\label{eq:tau_acc}
\end{align}
where $\tau^\prime_\mathrm{0,acc}$ is a constant
\citep[we adopt $\tau^\prime_\mathrm{0,acc}=1.61\times 10^8$\,yr appropriate for silicate, but the value for graphite has little difference;][]{2012MNRAS.422.1263H},
and $T_\mathrm{gas}$ is the gas temperature.

\subsubsection{Shattering and coagulation}\label{subsubsec:shattering}

Shattering and coagulation are described by grain--grain collisions
followed by the redistribution of fragments or coagulated grains 
\citep{1994ApJ...433..797J, 1996ApJ...469..740J,2009MNRAS.394.1061H}.
Shattering is assumed to occur only in the diffuse medium
($n_\mathrm{H}<1$\,cm$^{-3}$) where the
grain velocities are high enough. For coagulation, since we cannot spatially
resolve the dense and cold medium where it occurs, we adopt a sub-grid model.
We assume that coagulation occurs in gas particles which satisfy
$n_\mathrm{H}>10$\,cm$^{-3}$ and $T_\mathrm{gas}<1000$\,K,  and that
a mass fraction of $f_\mathrm{dense}=0.5$ 
of such a dense particle is condensed into
dense clouds, hosting coagulation, with $n_\mathrm{H}=10^3$\,cm$^{-3}$ and
$T_\mathrm{gas}=50$\,K on subgrid scales.

The time evolution of grain size distribution by shattering or coagulation
is expressed as
\begin{align}
\left[\frac{\partial\rho_\mathrm{d}(m,\, t)}{\partial t}\right]_\mathrm{shat/coag}
= -m\rho_\mathrm{d}(m,\, t)\int_0^\infty\alpha (m_1,\, m)\rho_\mathrm{d}(m_1,\, t)
\mathrm{d}m_1\nonumber\\
+ \int_0^\infty\int_0^\infty\alpha (m_1,\, m_2)\rho_\mathrm{d}(m_1,\, t)\rho_\mathrm{d}(m_2,\, t)
\mu_\mathrm{shat/coag}(m;\, m_1,\, m_2)\mathrm{d}m_1
\mathrm{d}m_2,\label{eq:shat}
\end{align}
where $\mu_\mathrm{shat/coag}$ describes the grain mass distribution function of
newly formed shattered fragments or coagulated grains, and
$\alpha$ is expressed in collisions between
grains with masses $m_1$ and
$m_2$ (radii $a_1$ and $a_2$, respectively) as
\begin{align}
\alpha (m_1,\, m_2)\equiv\frac{\sigma_{1,2}v_{1,2}}{m_1m_2},\label{eq:alpha}
\end{align}
where $\sigma_{1,2}=\upi (a_1+a_2)^2$ and $v_{1,2}$ are the collisional cross-section
and the relative velocity between the
two grains (we explain how to evaluate $v_{1,2}$ later).

We adopt the following formula for the grain velocity in a turbulent medium:
\begin{align}
v_\mathrm{gr}(a) &= 1.1\mathcal{M}^{3/2}\left(
\frac{a}{0.1~\micron}\right)^{1/2}\left(\frac{T_\mathrm{gas}}{10^4~\mathrm{K}}\right)^{1/4}
\left(\frac{n_\mathrm{H}}{1~\mathrm{cm}^{-3}}
\right)^{-1/4}\nonumber\\
&\times \left(\frac{s}{3.5~\mathrm{g~cm}^{-3}}\right)^{1/2}~\mathrm{km~s}^{-1},
\label{eq:vel}
\end{align}
where $\mathcal{M}$ is the Mach number of the largest-eddy velocity.
We adopt $\mathcal{M}=3$ for shattering and
$\mathcal{M}=1$ for coagulation to effectively achieve
the grain velocity levels suggested by
\cite{2004ApJ...616..895Y}.
In considering the collision rate between two grains
with $v_\mathrm{gr}=v_1$ and $v_2$, we estimate the relative velocity $v_{1,2}$ by
$v_{1,2}=(v_1^2+v_2^2-2v_1v_2\mu_{1,2})^{1/2}$,
where $\mu =\cos\theta$ ($\theta$ is an angle between the two grain velocities)
is randomly chosen between $-1$ and 1 in every calculation of $\alpha$.
The above expression for the velocity (with $\mathcal{M}=1$) was
originally derived by \cite{2009A&A...502..845O} 
who assumed the maximum eddy size to be 
the Jeans length and the maximum velocity
dispersion to be the sound velocity.
We simply use their functional form for the purpose
of avoiding grain velocity calculations in every time-step, but
modify it to scale the maximum velocity
(with the Mach number) to match with 
the velocity scale adopted for our previously
adopted models based on \cite{2004ApJ...616..895Y}.
As shown by \cite{2013EP&S...65.1083H}, 
detailed dependence of
grain velocities on the grain radius is not essential, but the overall velocity level is more important.
{
In particular, it is essential that large ($a\gtrsim 0.1~\micron$) grains attain
velocities higher than a few km s$^{-1}$ in the diffuse ISM, since
shattering, which plays an important role in the first production of small grains,
does not occur otherwise.
We will surely need to combine our simulation with direct dust motion calculations 
\citep[e.g.][]{2016MNRAS.456.4174H, 2018MNRAS.478.2851M}
to check if the velocity levels assumed here are actually achieved. Because of the
huge computational cost required, we leave this for a future work.}

For shattering, the total mass of the fragments is estimated following
\cite{2010Icar..206..735K}.
We consider a collision of two dust grains with masses $m_1$
and $m_2$. The total mass of the fragments ejected from $m_1$
is estimated as
\begin{align}
m_\mathrm{ej}=\frac{\phi}{1+\phi}m_1,
\end{align}
where
$\phi\equiv{E_\mathrm{imp}}/(m_1Q_\mathrm{D}^\star)$
[$E_\mathrm{imp}=\frac{1}{2}{m_1m_2}v_\mathrm{1,2}^2/(m_1+m_2)$
is the impact energy and $Q_\mathrm{D}^\star$
is the specific impact energy that causes the catastrophic disruption
(i.e.\ the disrupted mass is $m_1/2$)].
We adopt $Q_\mathrm{D}^\star =4.3\times 10^{10}$ cm$^2$ s$^{-2}$
(valid for silicate).\footnote{For graphite,
$Q_\mathrm{D}^\star =8.9\times 10^{9}$ cm$^2$ s$^{-2}$.}
Now we set the grain size distribution of shattered fragments.
The maximum and minimum grain masses of the fragments
are assumed to be
$m_\mathrm{f,max}=0.02m_\mathrm{ej}$ and
$m_\mathrm{f,min}=10^{-6}m_\mathrm{f,max}$, respectively
\citep[][]{2011A&A...527A.123G}.
We adopt the following fragment mass distribution including the remnant
of mass $m_1-m_\mathrm{ej}$ as
\begin{align}
\mu_\mathrm{shat}(m,\, m_1,\, m_2) &=
\frac{(4-\alpha_\mathrm{f})m_\mathrm{ej}m^{(-\alpha_\mathrm{f}+1)/3}}{3\left[
m_\mathrm{f,max}^\frac{4-\alpha_\mathrm{f}}{3}-
m_\mathrm{f,min}^\frac{4-\alpha_\mathrm{f}}{3}\right]}\,
\Phi (m;\, m_\mathrm{f,min},\, m_\mathrm{f,mmax})\nonumber\\
&+ (m_1-m_\mathrm{ej})\delta (m-m_1+m_\mathrm{ej}),\label{eq:frag}
\end{align}
where
$\Phi (m;\, m_\mathrm{f,min},\, m_\mathrm{f,mmax})=1$ if
$m_\mathrm{f,min}<m<m_\mathrm{f,mmax}$, and 0 otherwise,
$\delta $ is Dirac's delta function,
and $\alpha_{\rm f}$ is the power-law index of
the fragment size distribution
($\alpha_{\rm f} = 3.3$; \citealt{1996ApJ...469..740J}).
Grains whose radii are smaller than the minimum grain radius ($3\times 10^{-4}~\micron$)
are removed.

For coagulation, we adopt
\begin{align}
\mu_\mathrm{coag}=m_1\delta (m-m_1-m_2)
\end{align}
in equation (\ref{eq:shat}).

\subsection{Calculation of extinction curves}\label{subsec:ext}

Extinction curves describe the wavelength dependence of
the optical depth of dust.
Extinction curves have been useful in constraining the grain size
distribution \citep[e.g.][]{2001ApJ...548..296W}. Here, we calculate
extinction curves based on the grain size distributions, $n_{i}(a)$, where
$i$ indicates the composition of the dust grains 
(we consider multiple compositions as we explain below).
The extinction at wavelength $\lambda $ in units of magnitude 
is written as 
\begin{align} 
A_{\lambda}=(2.5 \log_{10}\mathrm{e})L\displaystyle\sum_{i}\displaystyle\int_{0}^{\infty}
n_{i}(a)\,\upi a^{2}Q_{\rm ext}(a, \lambda),
\end{align}
where $L$ is the path length, $Q_{\rm ext}(a, \lambda)$ is the extinction efficiency factor, which is evaluated
by using the Mie theory \citep[][]{1983asls.book.....B} 
with the same optical constants
for silicate and carbonaceous dust (graphite) as in
\cite{2001ApJ...548..296W}.
In this paper, the mass fractions of silicate and carbonaceous dust are fixed to 0.54
and 0.46, respectively (number ratio 0.43:0.57)
with both grain species having the same grain size distribution
\citep[][]{2009MNRAS.394.1061H}. 
Because this fraction is valid only for the Milky Way,
we concentrate on the comparison with the Milky Way extinction curve below.
For more comprehensive comparison, we need a more sophisticated model of
the evolution of grain compositions
\citep[][]{2016PASJ...68...94H},
which is left for a future work.
To concentrate on the extinction curve shape, the extinction is normalized to the value in the $V$ band
($\lambda ^{-1}=1.8\,\mu {\rm m}^{-1}$); that is, we output $A_\lambda /A_V$ (so that $L$ is cancelled out).

\section{Results}\label{sec:result}

\subsection{Evolution of grain size distribution}

\subsubsection{Dense and diffuse ISM}\label{subsubsec:dense_diffuse}

We show the general features in the evolution of grain size distribution.
This also serves as a test against the post-processing results in HA19.
We choose the gas particles contained in 
$0<R<7\,{\rm kpc}$ and $|z|<0.3\,{\rm kpc}$, where
we adopt a cylindrical coordinate with $R$ and $z$ being the
radial and vertical coordinates, respectively ($z=0$ is the disc plane).
In order to investigate the dependence on the local physical condition,
we divide the gas particles into two categories: 
the dense (cold) and diffuse (warm) gas.
The dense (cold) gas particles are chosen by $T_\mathrm{gas}< 10^{3}$\,K and
$n_\mathrm{H}> 10\,{\rm cm}^{-3}$, while the diffuse (warm) ones by
$0.1<n_\mathrm{H}<1$\,cm$^{-3}$ and $10^3 <T_\mathrm{gas} <10^4$\,K.
These criteria are the same as those adopted by HA19, 
but HA19 defined these by the physical condition at $t=10$\,Gyr. 
We use the physical
condition at each snapshot so that our definition in this paper really
traces the current dense and diffuse phases.
We choose the ages $t=0.3$, 1, and 3\,Gyr.
These ages are basically chosen for comparison with
A17. Although A17 also examined $t=10$\,Gyr snapshot,
the snapshot at $t=3$\,Gyr represents the
dust evolution at the oldest stage with a metallicity comparable to the present Milky Way;
thus, 
discussions in this paper do not change even if we choose $t>3$\,Gyr
instead of $t=3$\,Gyr.

\begin{figure}
	\includegraphics[width=\columnwidth]{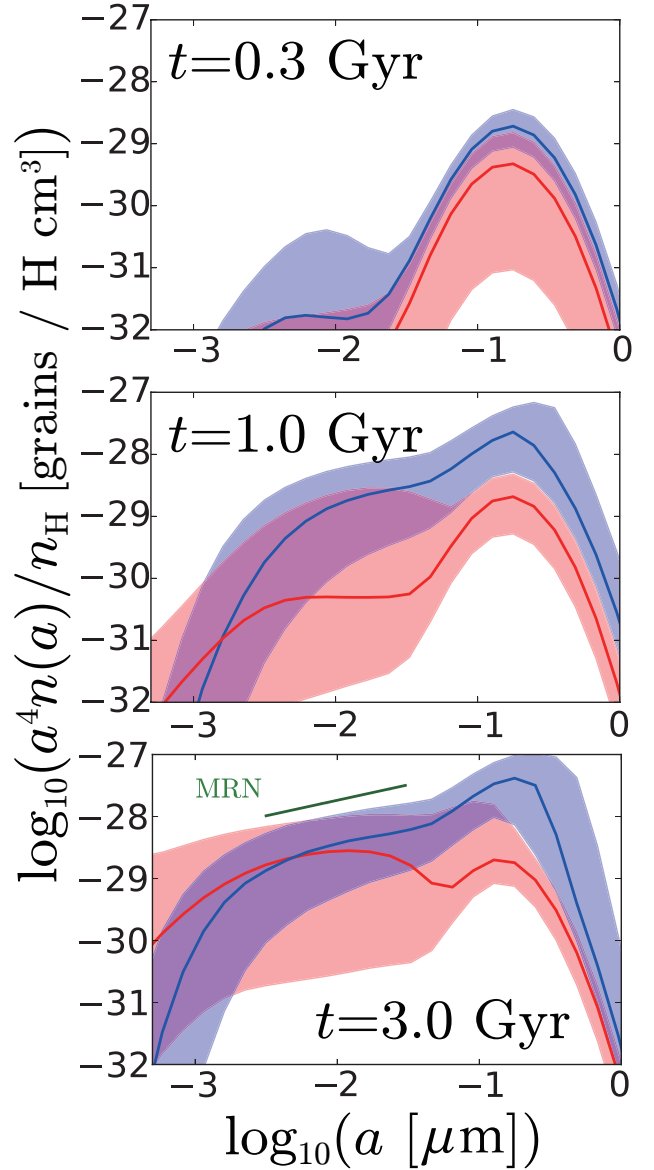}
        \caption{Evolution of grain size distribution for the entire galactic disc at $t=0.3$, 1, and 3 Gyr
        from top to bottom.
          The blue and red lines show the median grain size distribution in the dense and
          diffuse gas particles, respectively. The shaded regions show the areas
          between 25th and 75th percentiles.
          For the vertical axis, we present the grain size distribution per hydrogen (H)
          multiplied by $a^{4}$: the resulting quantity is proportional to
          the grain mass distribution per $\log_{10} a$. 
          The linear (green) solid line marked with `MRN' in the bottom panel shows
          the power-law slope of the MRN grain size distribution [$n(a)\propto a^{-3.5}$].
        }
          \label{fig:size}
\end{figure}

In Fig.~\ref{fig:size}, we show the grain size distributions at
each age for the dense and diffuse gas particles separately
with the median and 25th and 75th percentiles at each grain radius bin.
The dust is dominated by the large grains at young ages ($t\lesssim 0.3$ Gyr)
because stellar dust production is the major source of dust.
Since star formation tends to be associated with dense regions in
the galaxy, the peak in the grain size distribution is higher in the dense
medium than in the diffuse medium. This leads to more efficient small-grain
production by shattering in the dense gas particles because of higher rates of
grain--grain collisions.
In the dense ISM,
{ the small grains produced by shattering are further processed by accretion,
which forms a bump (second peak) at $a\sim 0.01~\micron$.
The bump is formed at smaller radii than the main peak because
the large surface-to-volume ratios of small grains lead to efficient accretion
\citep[][]{2012MNRAS.424L..34K}.}

As shown in the middle panel of Fig.~\ref{fig:size}
at $t=1$ Gyr, the grain abundance becomes larger at all grain radii
than at $t=0.3$ Gyr, because of further dust enrichment. The increase at
small grain radii is more significant than that at large radii because of
further dust processing by shattering and accretion. In particular,
dust and metal abundances are large enough for efficient accretion at this
stage. Therefore, the bump created by accretion
around $a\sim 0.01~\micron$ is commonly
observed in the grain size distributions. The dense gas particles have
more large grains than the diffuse ones for two reasons: one is the same as
above (we find more dense gas particles in actively star-forming regions),
and the other is more efficient coagulation in the dense ISM.
Indeed, we observe that coagulation depletes the smallest grains in the dense ISM.

In the bottom panel of Fig.~\ref{fig:size}, we show that at $t=3$ Gyr,
the evolutionary trend seen at $t=1$ Gyr is more pronounced.
Coagulation further proceeds especially in the dense medium,
producing a smooth power-law-like grain size distribution similar to the MRN
grain size distribution [$n(a)\propto a^{-3.5}$].
The effect of dust growth by accretion is also imprinted in the grain
size distribution in the diffuse ISM as seen in the bump around $a\sim 0.01~\micron$.
This imprinted bump is caused when these particles were previously included in
the dense ISM, {where accretion could take place efficiently if the
metallicity is high enough.}
Because of the variety in the past history, the diffuse
gas particles have larger variation in the grain size distribution especially at
small grain radii than the dense ones. In some regions as shown in
the lower 25th percentile, there is no bump at small grain radii, indicating that
accretion has not yet been efficient. Thus, we predict that there is a large
variety in the dust abundance in the diffuse ISM.

\subsubsection{Radial dependence}\label{subsubsec:radial}

\begin{figure}
	\includegraphics[width=\columnwidth]{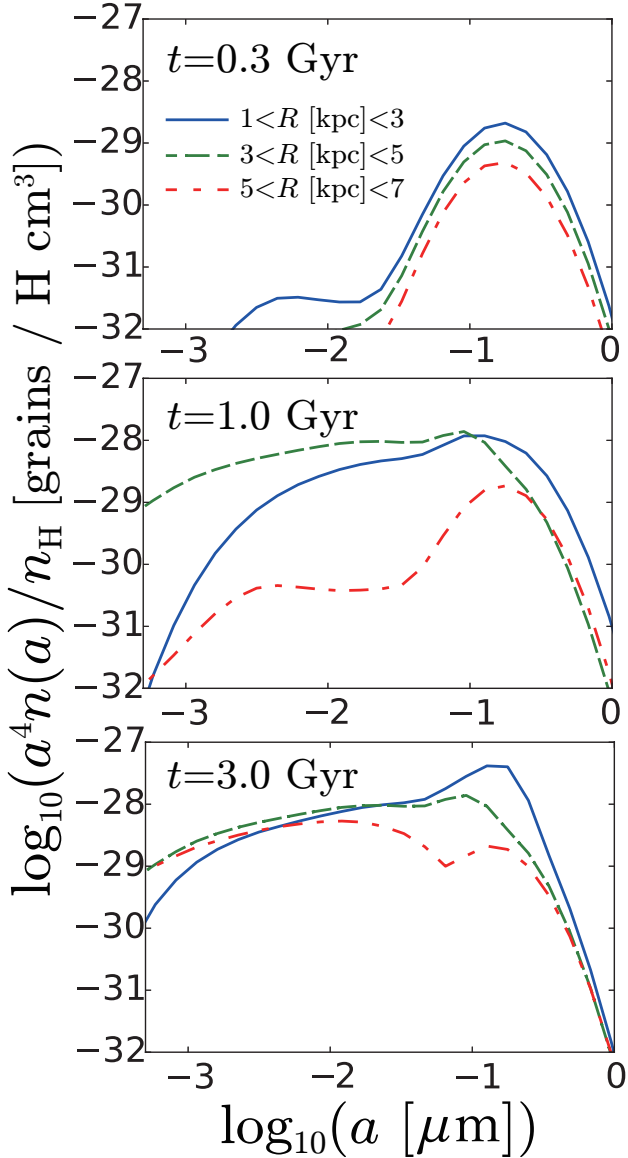}
        \caption{Grain size distribution at three different galactic radius ranges for the snapshots at $t=0.3, 1.0,$ 
        	and 3.0\,Gyr from top to bottom.
          The blue (solid), green (dashed) and red (dot-dashed) lines
          correspond to the radial ranges of $R=1$--3, 3--5, and 5--7\,kpc, respectively.
          In order to avoid the overlap of the shades, only median lines are displayed, but
          the dispersions are similar to those shown in Fig.~\ref{fig:size}.
    } \label{fig:sizeR}
\end{figure}

We analyze the spatial variation of grain size distribution.
In order to concentrate on the dust in the galactic disc, 
we select the gas particles in $|z|<0.3\,{\rm kpc}$.
In Fig.~\ref{fig:sizeR}, we show the radial dependence of
the grain size distribution in each snapshot by presenting the median 
at $R=1$--3, 3--5, and 5--7 kpc.
To make the presentation simple, we only plot the median, but
we expect similar dispersions to those shown in Fig.~\ref{fig:size}.
At $t=0.3$ Gyr, the grain size distributions in all regions are similar,
dominated by large grains produced by stars.
The dust abundance is the highest in the centre, which is due to
more active star formation and resulting metal enrichment
at smaller galactic radii.
Small-grain production by shattering is slightly seen
only in the inner region, 
but interstellar processing
is not efficient at all radii.

As shown in the middle panel of Fig.~\ref{fig:sizeR},
at $t=1.0$ Gyr, the grain size distributions are significantly
modified by interstellar processing. 
At the outer radii  ($R=5$--7 kpc),
the grain abundance is still dominated by large grains. At the intermediate
radii ($R=3$--5 kpc), accretion enhances the abundance of
small grains
because accretion time-scale of smaller grains is shorter than that of large ones
(Section \ref{subsubsec:dense_diffuse}).
The increased small grains also shatters
large grains, causing a sharp drop of grain size distribution
at $a\gtrsim 0.3~\micron$.
At the inner radii ($R=1$--3 kpc), in addition to the two processes above,
coagulation is also efficient because the grain abundance and gas density
are the highest among the three $R$ ranges. The effect of coagulation is clear in
the depletion of the smallest grains and the enhancement of the large grain
abundance at $a\gtrsim 0.1~\micron$. We note that the grain size distribution
does not change monotonically along the galactic radius: the strongest
enhancement of small grains relative to large grains is seen at
intermediate radii.

The bottom panel of Fig.~\ref{fig:sizeR} shows the grain size distributions
at $t=3$~Gyr. At this age, the small-grain abundance is enhanced to a similar
level at all radii.
In this phase, the initially observed log-normal shape is erased by interstellar processing.
The inner region has a higher large-grain abundance because of more efficient
coagulation. Since the inner region also hosts more efficient accretion, the dust
abundance is the highest.

\subsection{Evolution of dust abundance}\label{subsec:total}

\subsubsection{Dust-to-gas ratio and metallicity}

\begin{figure*}
	\includegraphics[width=0.95\textwidth]{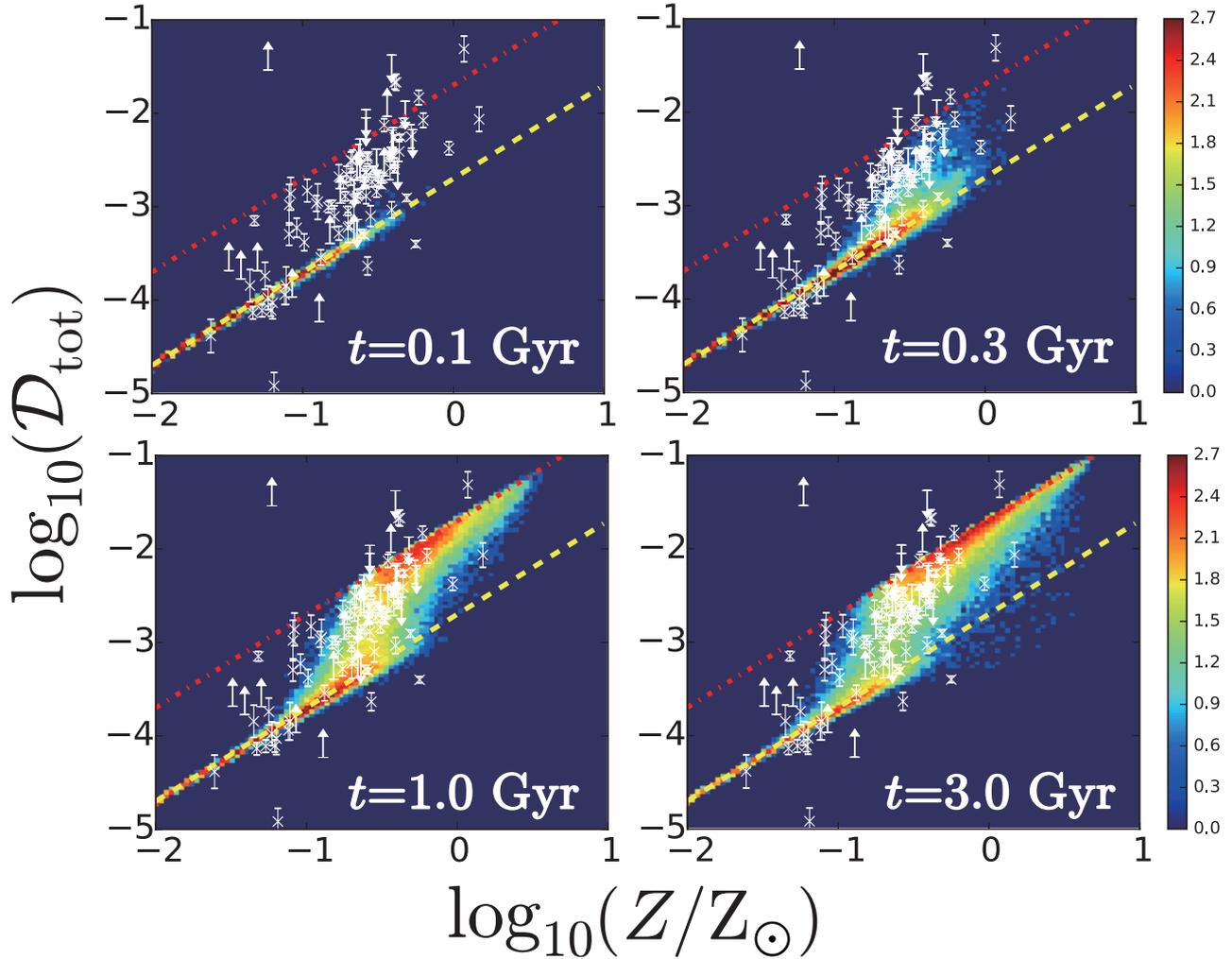}
        \caption{Distribution of gas particles 
          on the $\mathcal{D}_{\rm tot}$--$Z$ plane at 
          $t=0.1, 0.3, 1$ and 3 Gyr as labeled. 
          The colour indicates the logarithmic surface density of 
          the gas particle on this diagram as shown in the colour bars. 
          The yellow dashed and red dot-dashed lines present
          the linear relations of the stellar yield $(\mathcal{D}_{\rm tot}=f_{\rm in}Z)$
          and the saturation limit $(\mathcal{D}_{\rm tot}=Z)$, respectively.
          For comparison, we also show the observational data in \citet{2014A&A...563A..31R}
          and \citet{2014A&A...562A..76Z}
          (white points with error bars or upper/lower limits).
        }
    \label{fig:DZ}
\end{figure*}

We examine whether the evolution of the total dust abundance is reasonably
calculated by our model. The relation between dust-to-gas ratio
($\mathcal{D}_\mathrm{tot}$; equation \ref{eq:dg}) and metallicity ($Z$)
is often used to test if the dust evolution is correctly treated in a chemical
evolution framework 
\citep[e.g.][]{1998ApJ...496..145L}.
We plot the relation between $\mathcal{D}_\mathrm{tot}$ and
$Z$ for all gas particles in Fig.~\ref{fig:DZ}.
We also show a simple constant dust-to-metal ratios expected from
pure stellar dust production, $\mathcal{D}_{\rm tot}= f_{\rm in}Z$
(yellow dashed line), and 
the saturation limit, $\mathcal{D}_{\rm tot}= Z$ (red dot-dashed line).

We observe in Fig.~\ref{fig:DZ} that a fixed dust-to-metal ratio is 
a good approximation in the early phase ($t\la 0.1$~Gyr),  
since the dust evolution is driven by stellar sources.
Dust growth by accretion, whose efficiency (or time-scale)
has a metallicity dependence (equation \ref{eq:tau_acc}),
causes a nonlinear increase of $\mathcal{D}_\mathrm{tot}$
at $Z\gtrsim 0.1$ Z$_{\sun}$ as seen after $t=0.3$ Gyr.
Because of this nonlinearity, the dust-to-metal ratio is not constant
\citep[see also][]{1998ApJ...501..643D,1999ApJ...510L..99H,1999ApJ...522..220H,2003PASJ...55..901I,2008A&A...479..453Z,2013EP&S...65..213A}.
For example, \citet[][]{2003PASJ...55..901I} argued that 
the $\mathcal{D}_{\rm tot}$ increases rapidly by accretion when the galactic age is $\sim 0.3$\,Gyr.
Our simulation results are consistent with their results. 
As the system is enriched with metals, the $\mathcal{D}_{\rm tot}$--$Z$
relation extends towards higher metallicities and higher $\mathcal{D}_{\rm tot}$.
Because dust growth by accretion is limited by the available gas-phase metals
(Section \ref{subsubsec:destruction}),
the increase of $\mathcal{D}_{\rm tot}$ as a function of $Z$
becomes moderate at the highest metallicities. Naturally,
$\mathcal{D}_\mathrm{tot}$ does not exceed $Z$ (i.e.\ the simulation
data are always located below the line of
$\mathcal{D}_\mathrm{tot}=Z$).

We also plot the observed $\mathcal{D}_{\rm tot}$--$Z$
relation for nearby galaxies with each point corresponding to 
individual galaxy (not spatially resolved)
\citep{2014A&A...563A..31R,2014A&A...562A..76Z}. 
Note that
our plots show each gas particle within a single galaxy. Thus,
we use their observational data only as a first reference to judge if
our dust evolution model correctly reproduces the observationally expected
relation. Our model roughly reproduces the observation, 
indicating that our implementation of
various dust formation and processing mechanisms is successful in
catching the trend of dust evolution as a function of metallicity.

A17 also performed a similar hydrodynamic simulation of
an isolated galaxy with dust enrichment but using a simplified model for the grain size
distribution (two-size approximation). The $\mathcal{D}_{\rm tot}$--$Z$
relation in A17 (see their Fig.~7) is similar to that obtained in this paper.
Indeed, at $t\lesssim 0.3$ Gyr, the $\mathcal{D}_{\rm tot}$--$Z$
relation follows the one expected from the stellar dust production ($\mathcal{D}_{\rm tot}= f_{\rm in}Z$).
At $t\gtrsim 1$ Gyr, the non-linearity in the $\mathcal{D}_{\rm tot}$--$Z$
relation appears owing to dust growth by accretion, asymptotically approaching to
$\mathcal{D}_{\rm tot}=Z$. 
However, there are some differences between our results and A17's.
A17 showed a larger dispersion in the $\mathcal{D}_{\rm tot}$--$Z$ relation
even at $Z<0.1$ Z$_{\sun}$.
Since the increase of dust-to-gas ratio in this regime is driven by accretion (dust growth),
A17 shows a clear dust mass increase by accretion at $Z<0.1$ Z$_{\sun}$, while
our current simulation indicates that occurs accretion at $Z>0.1$ Z$_{\sun}$.
In other words, A17 has more efficient accretion.  Indeed, A17's fig.~8 shows that
the increase of small grains happens at $Z<0.1$ Z$_{\sun}$ at $t=0.3$ Gyr,
while our current simulation does not show any enhancement of small-grain abundance
at $t=0.3$\,Gyr (Fig.~\ref{fig:size}).
In the two-size approximation, all small grains are assumed to have a single radius of $5\times 10^{-3}~\micron$, while in our case shattering supplies a continuous spectrum of grain radii. Therefore, the accretion rate increases drastically once small grains are produced in the two-size approximation, while accretion is enhanced more gradually in our simulation. The difference indicates that there is a risk of overestimating dust growth by accretion at low metallicity in the two-size approximation.

\subsubsection{Radial dependence}\label{radial}

\begin{figure}
	\includegraphics[width=\columnwidth]{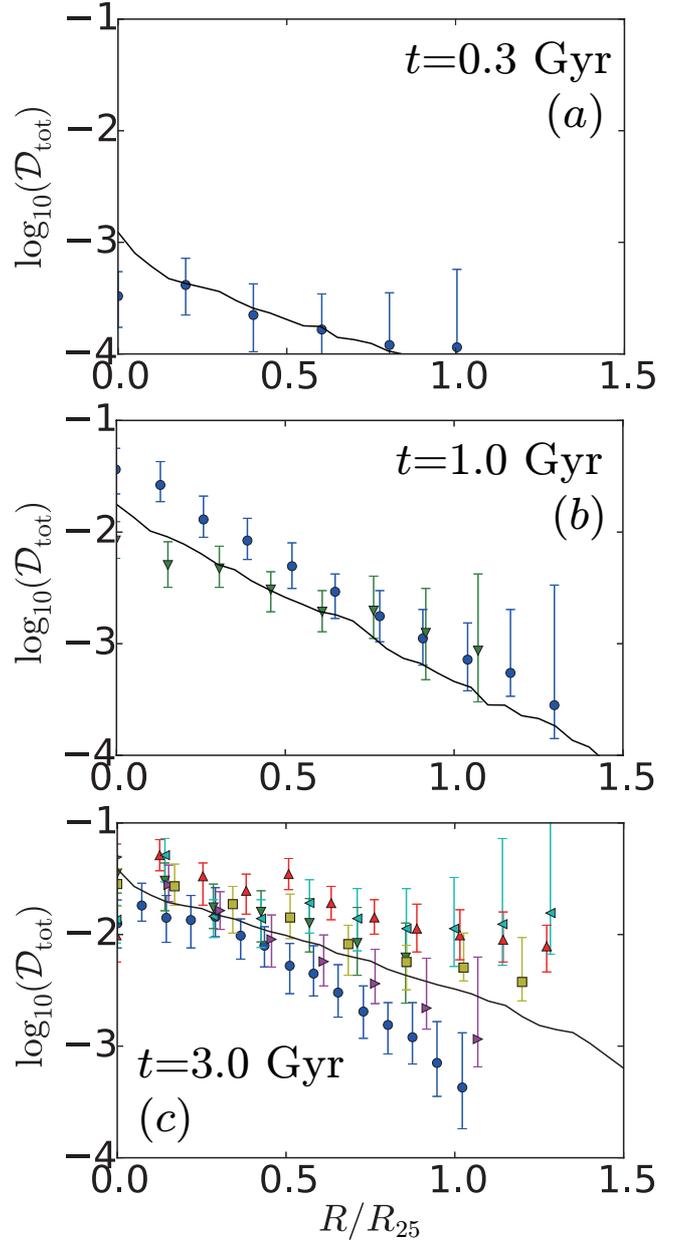}
        \caption{Comparison of the radial profiles of
          dust-to-gas ratio at $t=0.3$, 1, and 3\,Gyr
    [panels (\textit{a}), (\textit{b}), and (\textit{c}), respectively] with 
    the observational data {compiled by} \citet{2012MNRAS.423...38M}.
      The solid line is the radial profile of dust-to-gas ratio for each snapshot of the simulation.
      We scaled the radius with $R_{25}$ {(see text)}.
      In panel ({\it a}), the filled circles with error bars represent 
      the observational radial profile of Holmberg II. 
      In panel ({\it b}), the circle and triangle points show
      NGC 3621 and NGC 925, respectively.
      In panel ({\it c}), the circle, $\bigtriangledown $, $\bigtriangleup$,
      $\lhd $, $\rhd $, 
      and square symbols correspond to NGC 2403, NGC 4736, 
      NGC5055, NGC 5194, NGC628 and NGC 7793, respectively.
    }
    \label{fig:D_sequence}
\end{figure}

In Fig.~\ref{fig:D_sequence}, we compare the radial dependence of
the dust-to-gas mass ratio [here we express it as a function of $R$, 
$\mathcal{D}_{\rm tot}(R)$].
We adopt a mass-weight average for the dust-to-gas ratio as
\begin{align}
\mathcal{D}_{\rm tot}(R) \equiv \dfrac{\displaystyle\sum_{R-\Delta R/2 <R_i<R+\Delta R/2}
m^{\rm gas}_{i}\mathcal{D}^{\rm tot}_{i}}{\displaystyle\sum_{R-\Delta R/2 <R_i<R+\Delta R/2}m^\mathrm{gas}_i},
\label{eq:D_R}
\end{align}
where $R_i$ and $m^\mathrm{gas}_i$ are the radial coordinate and
the mass of the $i$th gas particle, respectively,
and the summation is taken
for all the gas particles in each radial bin of width $\Delta R$.
Here we choose gas particles in the disc
($|z|<0.3$\,kpc).
We use 40 bins on the linear scale in $0 \leq R \leq 10$ kpc. 

We adopt the same observational data as in A17.
{The data are
compiled by \citet{2012MNRAS.423...38M},
who originally adopted the sample from
\cite{2009ApJ...703.1569M} 
(chosen from the \textit{Spitzer} Infrared Nearby Galaxies Survey sample; \citealt{2003PASP..115..928K}).}
We adopt these data because of the completeness and uniformity of the quantities 
that are directly compared with the outputs
of our simulation, with a note that
there are some new data of spatially resolved dust emission
\citep{2019A&A...623A...5D,2018ApJ...865..117C}.
\cite{2012MNRAS.423...38M} investigated the radial profile of
dust-to-gas ratio and dust-to-metal ratio in nearby star-forming galaxies
(mainly spirals).  They used the metallicity calibration method 
of \citet[][]{2010ApJS..190..233M}.
We categorize the observational sample by typical ages using
the specific star formation rate (sSFR), following
the same criterion as in A17.\footnote{
  In this paper, we do not consider
  NGC 2841, NGC 3031, NGC 3198, NGC 3351, NGC 3521, and NGC 7331, 
  which are compared with the $t=10$ Gyr snapshot in A17 due to their
  low sSFRs ($<10^{-0.8}\,{\rm Gyr}^{-1}$).
  However, these galaxies have similar radial profiles as those used for the current 
  comparison at $t=3$ Gyr. Thus, even if we include the above galaxies into the comparison, 
  our current discussions do not change. }
We choose four ranges:  sSFR $> 10\,{\rm Gyr}^{-1}$,
$1\,{\rm Gyr}^{-1}<{\rm sSFR}< 10\,{\rm Gyr}^{-1}$ and sSFR $<1\,{\rm Gyr}^{-1}$,
which are referred to as Category I, II, and III, respectively.
They are compared with the snapshots at $t=0.3, 1$, and 3\,Gyr.
To cancel the galaxy size effect, we normalise
the radius by $R_{25}$ (the radius at which surface brightness falls to 25 mag arcsec$^{-2}$) 
for the observational sample, following \cite{2012MNRAS.423...38M}.
We also evaluate $R_{25}$ for the simulated galaxies 
by using the relation $R_{25}\simeq 4 R_{\rm d}$,
where $R_{\rm d}$ is the scale length of stellar disk
\citep{1998ggs..book.....E}. 
In order to obtain $R_{\rm d}$, we performed fitting for
the stellar surface density 
profile of simulated galaxy at each epoch.
The value of $R_{25}$ changes as time passes, because the distribution of
stellar component changes:
$R_{25} = 6.9$, 7.0, and 7.2\,kpc at $t = 0.3$, 1, and 3\,Gyr, respectively.

We plot the radial profile of $\mathcal{D}_{\rm tot}$ of each category
in Fig.~\ref{fig:D_sequence}.
The dust-to-gas ratio in the `youngest' phase represented by Category I
(Holmberg II) is broadly reproduced in terms of not only the slope
but also the absolute value (Fig.~\ref{fig:D_sequence}a).
In this phase, since the dust enrichment is dominated
by stellar dust production, the dust-to-gas ratio is proportional
to the metallicity; thus,
the variation of $\mathcal{D}_\mathrm{tot}$ along the galactic radius
directly traces the metallicity gradient.

In Fig.~\ref{fig:D_sequence}b, we compare
the galaxies in Category II with the simulation result at $t=1$\,Gyr. 
We observe that the level and slope of the radial profile are both reproduced well.
Compared with the result at $t=0.3$ Gyr, the dust-to-gas ratios are larger at all galactic
radii, and the slope is steeper. In the central part, accretion has drastically increased the
dust abundance, while stellar dust production is still dominant
at the outer radii.
Therefore, the contrast of dust-to-gas ratio between the central and outer parts is
the largest in this epoch.
Accordingly, the radial slope of $\mathcal{D}_\mathrm{tot}$ is the
steepest.

We further compare the galaxies in Category III with the simulation result 
at $t=3$\,Gyr in Fig.~\ref{fig:D_sequence}c. 
Although the observed data points have a large dispersion,
the simulation overall reproduces both the value and slope of $\mathcal{D}_\mathrm{tot}$.
In other words, the simulation result lies in the middle of the observational data.
Compared with the radial distribution at $t=1$\,Gyr, the dust-to-gas ratio is larger
at all radii, especially in the outer part, because dust growth by accretion has been
prevalent also there. The slope is shallower at $t=3$\,Gyr than at 1\,Gyr.

The dust-to-metal ratio traces the fraction of metals condensed into the solid phase.
Different processes affect the dust-to-metal ratio differently.
{Indeed, \cite{2012MNRAS.423...26M} showed by analytic
arguments that the
radial profile of dust-to-metal
ratio could be used to identify the dominant dust enrichment process.} 
The dust-to-metal ratio in each radial bin is estimated by dividing the mass-weighted
dust-to-gas ratio by the mass-weighted metallicity as
\begin{align}
  \dfrac{\mathcal{D}_{\rm tot}}{Z}(R)=
  \dfrac{\displaystyle\sum_{R-\Delta R/2 <R_i<R+\Delta R/2}m^{\rm gas}_{i}D_{{\rm tot},i}}{\displaystyle\sum_{R-\Delta R/2 <R_i<R+\Delta R/2}m^{\rm gas}_{i}Z^{\rm gas}_{i}}\, ,
\end{align}
where the summation is taken in each radial bin in the same way as in equation (\ref{eq:D_R})
with a constraint of $|z|<0.3$\,{kpc}.

\begin{figure}
	\includegraphics[width=\columnwidth]{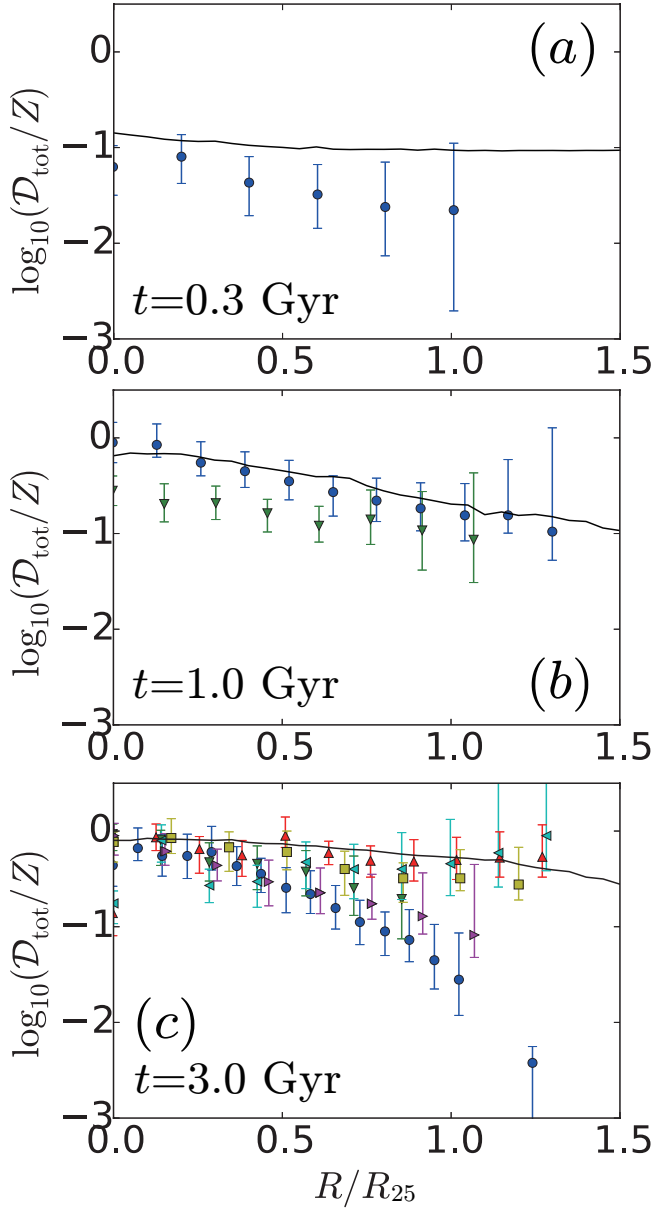}
        \caption{
          Same as Fig.~\ref{fig:D_sequence} but presenting the radial profile of
          dust-to-metal ratio. 
        }
    \label{fig:DZ_sequence}
\end{figure}

In Fig.~\ref{fig:DZ_sequence},
we find that the dust-to-metal ratio is almost constant in the early phase at $t=0.3$\,Gyr
and that the flat trend is broadly consistent with the data of Holmberg II (Category I)
considering the large errors.
At this stage, the level of dust-to-metal ratio is broadly determined 
by the condensation efficiency in stellar ejecta. 
Thus, the dust-to-metal ratio is almost constant ($\simeq f_{\rm in} = 0.1$),
and its radial profile is flat.

For the galaxies in Category II, which we compare with the simulation at
$t=1$\,Gyr in Fig.~\ref{fig:DZ_sequence}b, 
we find that the profile of dust-to-metal ratio is broadly consistent with the
observational data, considering the large error bars.
The decreasing trend of dust-to-metal ratio with increasing radius 
is caused by more efficient accretion in the centre than in the outer regions, 
since accretion favours metal-rich environments
\citep[see also][]{2012MNRAS.423...26M,2012MNRAS.423...38M}.
Accordingly, the slope of the dust-to-metal ratio profile is the steepest around this epoch.
We emphasize that, since the dust-to-metal ratio is not constant, we should at least
take non-constant dust-to-metal ratios (both temporally and spatially)
into account in modeling the dust evolution in galaxies.

The galaxies in Category III are compared with the snapshot at $t=3$\,Gyr
in Fig.~\ref{fig:DZ_sequence}c. 
The simulation somewhat overproduces the dust-to-metal ratios at large radii, 
although some galaxies are consistent with our result even at the outer radii.
The radial gradient of dust-to-metal ratio is shallower at $t=3$\,Gyr than at $t=1$\,Gyr, 
because the effect of accretion becomes more prevalent up to larger radii at later times.
NGC 2403 shows a steep decline of dust-to-metal ratio at the outer radii, 
which cannot be explained by our models.

The radial profiles of dust-to-gas ratio and dust-to-metal ratio are 
similar to the previous simulation that used the two-size approximation (A17).
This means that the two-size approximation
adopted in A17 is good enough if we only calculate the total dust abundance.
We point out that the tendency of overestimating the dust-to-metal ratio
at $t=3$\,Gyr
(Fig.~\ref{fig:DZ_sequence}c) is also observed in A17. This could be due to an overestimate
of dust growth by accretion. We discuss this issue in Section~\ref{sec:discussion}.

\subsection{Extinction curves}

\subsubsection{Dependence on the ISM phases}

\begin{figure*}
	\includegraphics[width=0.9\textwidth]{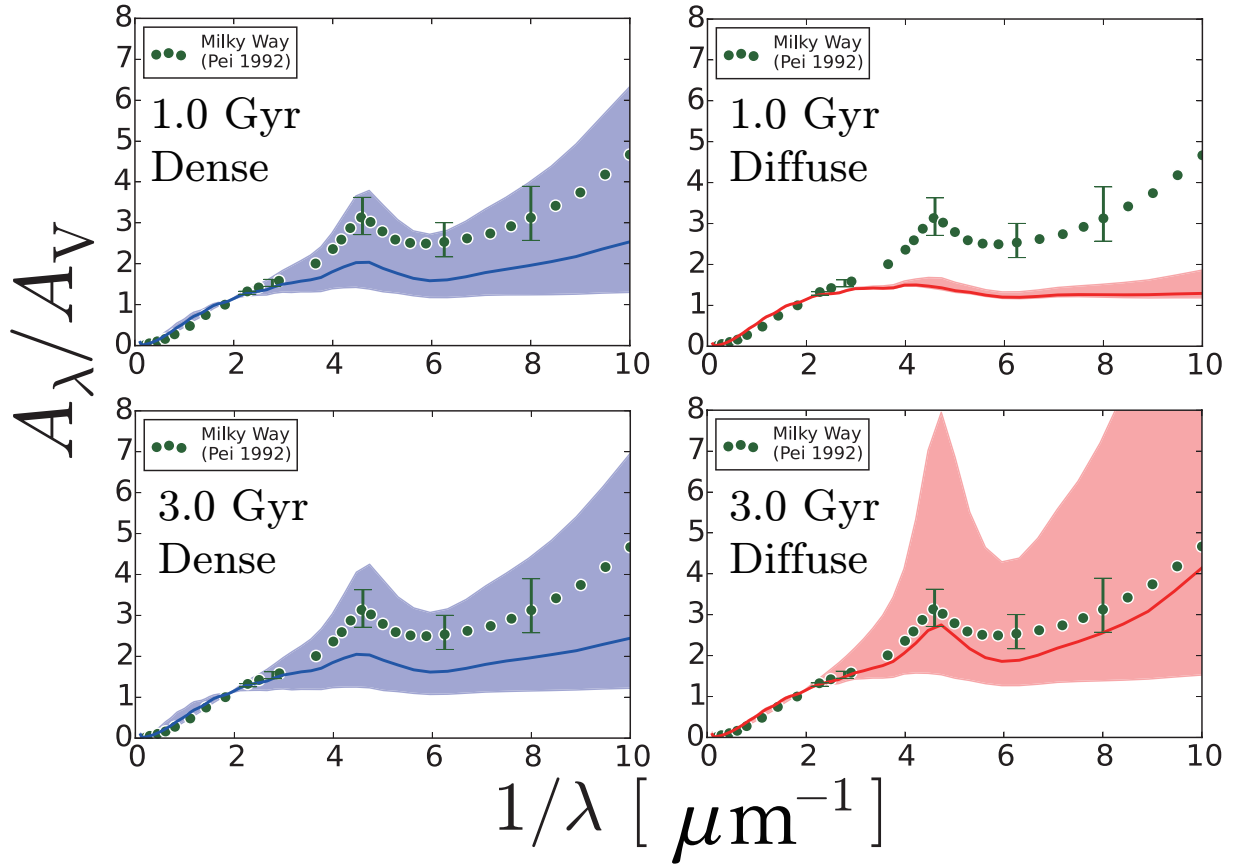}
        \caption{
          Variation of extinction curves at $t=1$ and 3\,Gyr for the dense and 
          diffuse gas particles as indicated in each panel. 
          The solid line is the median and the shaded region
          shows the area between the 25 and 75 percentiles.
          We also show the observed Milky Way extinction curve \citep[][]{1992ApJ...395..130P} by
          the green circles. The error bars show the
          dispersion for various lines of sight
          taken from \citet{2007ApJ...663..320F} \citep[see also][]{2013ApJ...770...27N}.
        }
    \label{fig:Ex}
\end{figure*}

As representative observable dust properties
that reflect the grain size distribution,
we calculate the extinction curves by the method described
in Section~\ref{subsec:ext}.
In Fig.~\ref{fig:Ex}, we show the extinction curves
for the dense and diffuse gas particles.
We sorted the extinction curves ($A_\lambda /A_{\rm V}$) and 
obtained the median and 25th and 75th percentile
for each of the dense and diffuse particles. 
We only show the extinction curves at $t = 1$ and 3\,Gyr.
At $t<1$\,Gyr, the extinction curves are similar to the ones in the diffuse gas at $t=1$\,Gyr,
because they are determined by the log-normal grain size distribution
of stellar dust. Thus, we only focus on $t\ge 1$\,Gyr.
At $t = 1$\,Gyr, the extinction curves in the diffuse medium are flat as mentioned above.
In the dense regions, in contrast, the extinction curves
have more variety, reflecting more efficient small grain production as shown in Fig.~\ref{fig:size}.
As explained in Section~\ref{subsubsec:dense_diffuse}, the dust-to-gas ratio in the dense
gas particles tends to be higher since the dense regions are more metal-enriched. 
Thus, shattering and accretion affect the grain size distributions 
more in the dense medium than in the diffuse medium. 
Consequently, steeper extinction curves appear in the dense ISM at $t=1$\,Gyr.

At $t=3$\,Gyr, the extinction curve shape has a large variety at
UV wavelengths, especially in the diffuse ISM.
As we observe in Fig.~\ref{fig:size}, the variety
in the grain size distribution is large at grain radii $a\lesssim 0.03\,\micron$ at
$t = 3$\,Gyr, which affects the extinction at $\lambda \sim 2 \pi a \sim 0.2~\micron$.
Since the diffuse gas particles have a larger variety in the grain size distribution
around $a\sim 0.03\,\micron$ than the dense ones, 
the variety in the extinction curves is greater in the diffuse ISM. 
The median extinction curve is slightly
steeper in the diffuse ISM at $t=3$\,Gyr, contrary to the results at $t=1$\,Gyr.

In Fig.~\ref{fig:Ex}, we also plot the observed Milky Way extinction curve as a
reference. The observational data is taken from
\cite{1992ApJ...395..130P} and the dispersion adopted from
\citet{2007ApJ...663..320F} and \citet{2013ApJ...770...27N} is also shown at
some wavelengths.
Since our extinction curve model adopts the often used dust
properties well calibrated by the Milky Way condition, 
we only compare our extinction curves with the Milky Way curve.
{Also, our simulation targets a Milky-Way-like (spiral) galaxy, so it would make sense
to focus on the Milky Way extinction curve.
Including the variation of dust composition needs
a more sophisticated evolution model for the grain size distribution.
The extinction curves of the Large and Small Magellanic Clouds could also be reproduced by changing the ratio of silicate and graphite with the same grain size distribution \citep[][]{1992ApJ...395..130P}.
More comprehensive modelling
of extinction curves is left for a future work.}

We observe that the median in the diffuse medium at $t=3$\,Gyr
broadly explains the Milky Way extinction curve.  
The dispersion in the diffuse ISM at $t=3$\,Gyr is much larger
than the observed one. 
Note that the dispersion among the individual gas particles is calculated
in our model, while an observed extinction curve
{is an integrated quantity} 
along a line of sight.
Therefore, it may be natural that the actually observed extinction curves show
a much smaller dispersion than our particle-based statistics.
We also find that the extinction curves in the dense ISM at $t=3$\,Gyr
is flatter
than the Milky Way extinction curve. This is due to coagulation.
On the other hand, the extinction curves at $t=1$\,Gyr in the dense ISM
is steeper than those
at $t=3$\,Gyr, and the dispersion covers
the observed Milky Way extinction curve.
This steepness of extinction curve is due to efficient dust growth by accretion.
Thus, our model is capable of reproducing the Milky Way extinction curve
at $t\gtrsim 1$\,Gyr.

The difference in the extinction curves between the diffuse and dense ISM is prominent at
each age, which indicates the importance of hydrodynamical evolution on the grain
properties. Also, it is interesting that the trend is reversed between $t=1$ and 3\,Gyr:
the extinction curves are steeper in the dense ISM at an early age, while they are flatter
in the dense ISM at a later stage. This trend is consistent with our previous calculation
based on the two-size approximation \citep{2017MNRAS.469..870H}. 
\cite{2014MNRAS.437.1636H} also reproduced the correlation
between the bump strength and steepness of extinction curves in the Milky Way
by coagulation in the dense ISM. 
In this scenario, extinction curves are flatter in the dense ISM,
which is consistent with our results in the later age.

\subsubsection{Radial dependence}

\begin{figure*}
	\includegraphics[width=0.9\textwidth]{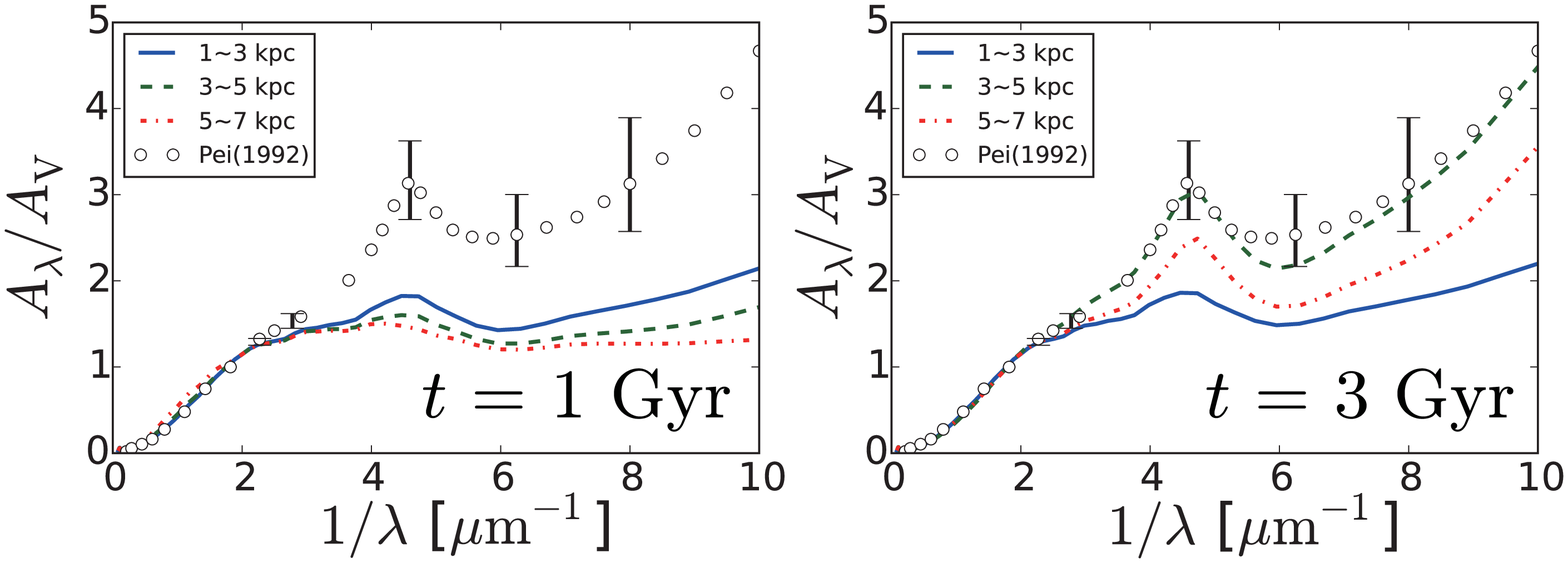}
        \caption{
          Radial dependence of extinction curves at $t=1$ and 3\,Gyr.
          The blue (solid), green (dashed) and red (dot-dashed) lines
          correspond to the radial range of $R=1$--3, 3--5, and 5--7\,kpc, 
          respectively. We also compare the calculated extinction curves with the
          same observational data as in Fig.~\ref{fig:Ex}.     }
    \label{Fig:extinction_radii}
\end{figure*}

We investigate the dependence of extinction curves on the galactic radius
in a way similar to \cite{2017MNRAS.469..870H}, who performed the analyses
based on our previous simulation with the two-size approximation.
Following their paper, we adopt
a galactic-radius bin width of 2\,kpc and take the mass-weighted 
average for all gas particles in each radius bin.
In order to focus on the extinction curves in the disk region,
we exclude the central 1\,kpc region. 
Indeed, the metallicity in the central 1\,kpc region
is relatively high $Z\gtrsim 3$ Z$_{\sun}$ \citep[see in fig.\ 6 of][]{2019MNRAS.484.2632S}
and HA19 also excluded the very central region ($R<0.1$\,kpc) for estimating the extinction curves.
We analyse the extinction curves up to $R = 7$\,kpc.
Fig.~\ref{Fig:extinction_radii} shows the extinction curves in 
the three radial bins at $t=1$ and 3 Gyr.

We observe in Fig.~\ref{Fig:extinction_radii} that
the features of extinction curves (the 2175 \AA\ bump
and the UV slope) become more prominent with age,
which is caused by the increase of small grains relative to large grains.
Small grains are more efficiently produced by shattering and
accretion at small $R$ in the early epoch as mentioned in Section~\ref{radial}.
As the dust-rich regions extends to the outer radii with age,
shattering and accretion become efficient there; thus, the outer regions
also have steep extinction curves at $t=3$\,Gyr. We give more
detailed descriptions in what follows.

At $t=1$\,Gyr, the extinction curves at $R<3\,{\rm kpc}$ 
show a prominent 2175 \AA\ bump and UV rise.
The extinction curves at $R > 3$\,kpc are still flat
because the dust abundance is dominated by large grains produced by stars.
The radial variation of extinction curves at 3\,Gyr shows the opposite trend to that at 1\,Gyr, 
with the outer regions ($R>3$ kpc) having steeper extinction curves 
compared with the inner regions ($R<$ 3 kpc). 
In this epoch, accretion dominates the dust evolution in the entire galaxy regions as mentioned
above. Therefore, extinction curves have steep slopes in the UV.
As discussed in Section~\ref{subsubsec:radial}, 
coagulation becomes efficient in the central part, so that the
extinction curves are flattened at $R<3$\,kpc.
The extinction curves are the steepest in the intermediate radii ($R\sim 3$--5\,kpc)
since accretion is efficient but coagulation is not yet active there. At larger radii
($R>5$\,kpc), the extinction curves are flatter because of less efficient accretion.

\section{Discussion}\label{sec:discussion}

\subsection{Total dust abundance}

Although the main focus of this paper is the grain size distribution, 
the model should reproduce the evolution of the total dust abundance. 
In Section~\ref{subsec:total},
we have shown that our simulation reproduces the total dust abundance (dust-to-gas ratio)
in nearby galaxies. 
The relation between dust-to-gas ratio and metallicity and the radial profiles of
dust-to-gas ratio and dust-to-metal ratio are broadly consistent 
with the observational data of nearby galaxies.

However, there are some improvements necessary for better agreement with the observational
data. In Section~\ref{radial}, we find that the radial profile of dust-to-metal ratio
tends to be above the observational data points,
although it is not completely discrepant from the observational data. 
This raises the following discussions in the early and late epochs.
In the early epoch at $t=0.3$\,Gyr, the possible discrepancy is due to the
overestimate of dust condensation efficiency in stellar ejecta, for which we adopted
$f_\mathrm{in}=0.1$.   
In fact, the condensation efficiency
is still uncertain: for SNe, the dust mass injected to the ISM is
strongly affected by the reverse shock destruction, 
which is sensitive to the density of the ambient medium
\citep{2007MNRAS.378..973B,2007ApJ...666..955N}.
For the dust yield in AGB stars, there are still some variety among
different condensation calculations
as compiled by \citet{2011EP&S...63.1027I} and \citet{2013MNRAS.436.1238K}.
Considering such uncertainties, the agreement within a factor of 2 is good enough.
Moreover, the almost flat slope is consistent with the observational data, 
which suggests that the stellar dust production is dominant at all galactic radii.

At $t\gtrsim 1$\,Gyr, the radial profile of dust-to-metal ratio has a negative
gradient, because dust growth by accretion starts to play a role from the central part.
The existence of such a negative gradient is consistent with the observational data.
However, the calculated radial profiles may be too flat compared with the observational
data (see Fig.~\ref{fig:DZ_sequence}).
The flat radial profile is due to efficient dust growth by accretion up to the large galactic radii;
thus, accretion is probably too efficient in our calculation. 
There are some possible improvements for the treatment of accretion. 
We equally treated all metal elements but dust has some specific composition. 
Ideally, we should model accretion so that it reproduces the actual dust
composition (such as silicate).
\cite{2017MNRAS.469..870H} treated two grain species (silicate and carbonaceous dust)
separately by neglecting a formation of compound species. 
This is also a simplification,
and it may underestimate the accretion efficiency 
because each grain species accretes only specific elements. 
Separating the grain species is also important for the prediction 
of extinction curves. 
In the future, we will work on a framework that separates the grain species.

\subsection{Robustness and uncertainty in the grain size distribution}

The shape of grain size distribution is determined by the stellar dust production
in the early (or low-metallicity) phase of galaxy evolution.
Thus, as long as stars predominantly produce large ($a\gtrsim 0.1\,\micron$)
grains, the resulting grain size distribution and extinction curve are robust.
As mentioned in the Introduction, most theoretical calculations of dust condensation
in stellar ejecta indicate that such large grains are favourably produced.

The grain size distribution in later epochs, on the other hand, is determined
by interstellar processing.
{ In our model, shattering plays an important role 
in the initial
production of small grains. As mentioned in
Section \ref{subsubsec:shattering}, it is essential that the largest grains have
velocities { high} enough for shattering ($\gtrsim$ a few km s$^{-1}$) in the diffuse ISM.
Otherwise, the { initial} production of small grains does not occur efficiently. Subsequently,}
accretion and coagulation are important
in increasing small and large grains, respectively. The final shape of the grain
size distribution approaches the one similar to the MRN grain size distribution.
Collisional processes such as coagulation are shown to robustly lead to a grain size distribution
like MRN 
\citep[e.g.][]{1969JGR....74.2531D,1994Icar..107..117W,1996Icar..123..450T,2010Icar..206..735K}.
The upper limit of the grain radii is rather sensitive to the subgrid model of
coagulation as shown in HA19. If the maximum grain radius becomes as large as
$\sim 1\,\micron$, the extinction curve becomes too flat
to explain the Milky Way curve.
Although $\micron$-sized grains should exist in some dense molecular
cloud cores \citep[][]{2010Sci...329.1622P,2014A&A...572A..20L},
such large grains should not be the majority of the ISM
\citep[see also][]{2009SSRv..143..333D}.
To determine the maximum grain size without any subgrid model, it is
necessary to spatially resolve dense molecular cloud cores. This is challenging
at this moment, because the simulation box should be wide enough to cover
a representative region of a galaxy (i.e.\ $\sim$kpc scale) 
but it should still resolve the sub-pc scales.
Nevertheless, we could argue that our subgrid model is successful because the maximum
grain radius is consistent with MRN ($\sim 0.25~\micron$) at $t=3$~Gyr
(Fig.~\ref{fig:size}).

\subsection{Importance of implementing dust evolution in hydrodynamic simulations}

As mentioned above, direct implementation of dust evolution calculations in
hydrodynamic simulations is important since some grain evolution processes occur on
short time-scales.
According to equation (\ref{eq:tau_acc}), accretion occurs on a time-scale
shorter than $10^{7}$ yr for small grains in solar-metallicity environments.
The post-processing analysis by HA19 was based on snapshots with 
intervals of $\sim 10^7$ yr, which is too long to capture the change of physical conditions
(density and temperature) for accretion. Coagulation also occurs on
a comparable time-scale in a solar-metallicity environment.
Thus, it is essential to solve the dust evolution and hydrodynamics 
simultaneously and consistently.

In our simulation, individual SNe cannot be resolved; 
thus, the rate of SNe affecting
a certain gas particle ($\gamma$ in equation \ref{eq:tau_dest}) is
used to estimate the dust destruction rate at each time step. 
Since SN destruction counteracts the increase of
dust mass by accretion which could also have a short time-scale,
post-processing does not give precise results.
Moreover, the physical condition of the ISM changes rapidly 
when the SN energy is deposited, and the lifetime of superbubbles is
as short as $\sim 10^6$\,yr \citep[e.g.][]{2011piim.book.....D}.
To correctly trace the dust processing induced by the energy input
from SNe, we need to solve the grain evolution and hydrodynamics simultaneously.

\subsection{Dependence on the subgrid treatment}

\begin{figure*}
	\includegraphics[width=0.9\textwidth]{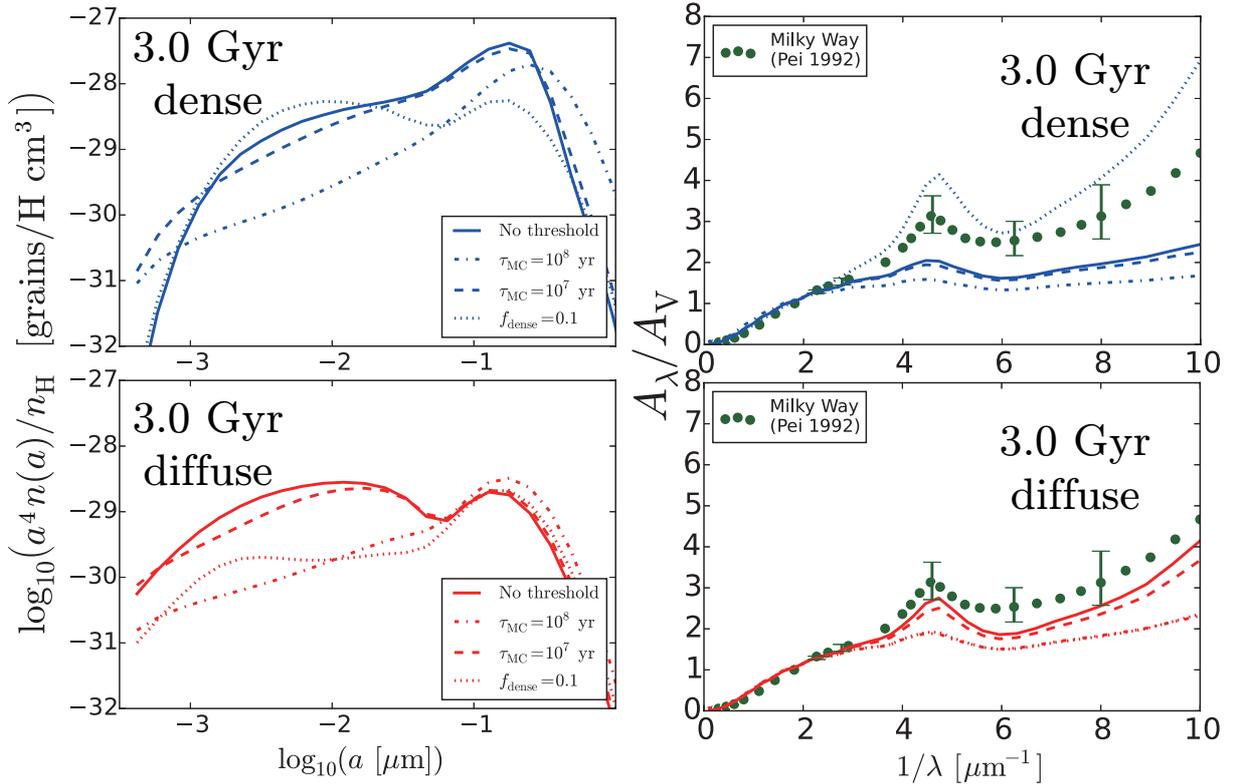}
        \caption{
          Some additional tests for the subgrid treatment of accretion and coagulation.
          We show the grain size distributions (left) and extinction curves (right) in the dense (upper) and
          diffuse (lower) gas particles at $t=3$\,Gyr.
          We only show the medians.
          The solid line shows our `default' simulation, which is
          the same as in Figs.\,\ref{fig:size} and \ref{fig:Ex}.
            The dashed and dot--dashed lines correspond to the cases with molecular cloud lifetime of $\tau_\mathrm{MC}=10^{7}$ and $10^{8}$ yr, respectively (see text).
            The dotted lines represent the case of a different dust condensation efficiency $f_{\rm dense}=0.1$ without imposing the lifetime of molecular clouds. 
            In the right-bottom panel (extinction curves), the dot--dashed and dotted lines
            overlap with each other.
          }
    \label{Fig:MC}
\end{figure*}

{ In our simulations, the dust processing in dense clouds 
(accretion and coagulation) is treated by subgrid modelling, 
because} the typical size of dense clouds (or molecular clouds) 
is much smaller than the typical softening length of gas particles ($\sim$80 pc). 
Therefore, it is worth examining the robustness of the results by performing additional simulations.
  We first examine the dependence on $f_\mathrm{dense}$ by lowering it to 0.1
  (the fiducial value is 0.5). We also test the effect of molecular cloud lifetime,
  which is determined by the gas dynamical evolution on subgrid scales.
  We only show the results at $t=3$\,Gyr, when the effects of accretion and coagulation
  appear most significantly among the ages chosen above.
 
First, we show the results with $f_\mathrm{dense}=0.1$ in Fig.\ \ref{Fig:MC}.
We only show the medians.
 We observe that, compared with the fiducial case with $f_\mathrm{dense}=0.5$,
 the small-grain abundance is enhanced while the large-grain abundance is suppressed
 in the dense regions. The grain size distribution stays bimodal because coagulation is too weak
 to create a smooth grain size distribution. Accordingly, the extinction curves become steeper
 in the dense medium.
 { In the diffuse medium, on the other hand, the abundance of grains with
 $a\sim 0.01~\micron$ is less for the case of smaller $f_\mathrm{dense}$.}
 Accordingly, the extinction curves are flat in the diffuse ISM.
  Recalling that accretion and coagulation dominate the shaping of grain size distribution
  at a later stage, low $f_\mathrm{dense}$ indicates that the evolution of grain size distribution
  occurs slowly. This explains the difference between $f_\mathrm{dense}=0.5$ and 0.1. 

Next, we examine the effect of molecular cloud lifetime.
We implicitly assumed that the mass exchange between
molecular clouds and other gas is frequent enough so that the
uniform grain size distribution is achieved on the subgrid scale.
However, if such an exchange
does not occur, dust growth by accretion could be saturated
in each molecular cloud. In this case, accretion is regulated by the lifetime
of molecular clouds rather than the actual accretion time-scale
\citep[][]{2000PASJ...52..585H}. 
To examine this effect, we set $\tau_\mathrm{acc}=\tau_\mathrm{MC}$
\textit{if} $\tau_\mathrm{acc}$
(equation \ref{eq:tau_acc}) is shorter
than $\tau_\mathrm{MC}$ (molecular cloud lifetime).
The lifetime of molecular clouds are uncertain, but it is likely to be on
  the order of $10^{7}$ yr \citep{2006MNRAS.369.1201W,2007ARA&A..45..339B}.
  \cite{2009ApJ...700L.132K} suggested longer lifetimes ($\sim 10^8$ yr) based on the
  observed stability of dense structures. Thus, we examine $\tau_\mathrm{MC}=10^7$ yr and
  $10^8$ yr.
  { Since $\tau_\mathrm{acc}$ estimated by equation (\ref{eq:tau_acc}) is
  inversely proportional to the
  metallicity, $\tau_\mathrm{acc}=\tau_\mathrm{MC}$ is adopted only if the metallicity is high enough.}
  Coagulation has no saturation; therefore 
  we only change the prescription for accretion here.
 
The result for $\tau_\mathrm{MC}=10^7$ yr
is shown in Fig.\,\ref{Fig:MC}.
We observe that, if $\tau_\mathrm{MC}=10^7$ yr, the results do not change much
for the grain size distribution and extinction curves.
This means that the accretion time-scale is mostly longer than $10^7$\,yr
{ (i.e.\ $\tau_\mathrm{acc}>\tau_\mathrm{MC}$ in most of the cases so that
$\tau_\mathrm{acc}=\tau_\mathrm{MC}$ is rarely adopted)}.
In contrast, if $\tau_\mathrm{MC}=10^8$\,yr,
the abundance of grains at
$a\lesssim 10^{-1.5} \,\mu {\rm m}$ is significantly suppressed. 
{ This is because of less efficient accretion.
If $\tau_\mathrm{MC}$ is as long as $10^8$ yr, $\tau_\mathrm{acc}$ estimated by
equation (\ref{eq:tau_acc}) more often becomes shorter than $\tau_\mathrm{MC}$
(i.e.\ $\tau_\mathrm{acc}=\tau_\mathrm{MC}$ is more often adopted).}
  We also observe that the abundance of the biggest grains is higher for this case because
  large grains are shattered less by small grains. Because of the dominance of large grains,
  extinction curves are very flat in both dense and diffuse media.

In summary, the subgrid treatment of accretion and coagulation 
affects the results.
{ The fraction of dense gas (or the structure of gas density
below a few ${}\times 10$ pc) and the molecular cloud lifetime are particularly important.
It is desirable to determine or constrain these quantities by higher-resolution
simulations and/or observations.
}

\subsection{Future prospect}
In this paper, we only simulated an isolated disc galaxy.
Although we expect that this work gives a representative case for dust evolution,
there are clearly other types of galaxies with various star formation time-scales
and mass scales.
To calculate the evolution of various types of galaxies,
cosmological simulations would be ideal. Therefore, a natural extension of
this work is to implement the evolution of full grain size distribution in cosmological
hydrodynamic simulations. Since cosmological calculations have worse spatial
resolution, the calculation results in this paper will give a useful calibration
for similar galaxies produced in a cosmological box.

Cosmological zoom-in simulations that focus on some specific halos 
are also useful to achieve a high spatial resolution. 
There are some examples of such simulations using zoom-in methods
\citep{2015MNRAS.451..418Y,2016MNRAS.457.3775M} which revealed
the dust distribution in galaxies. In particular, high spatial resolution is important
to predict detailed radiative properties since 
the wavelength dependence of dust absorption at UV and optical wavelengths
in the entire galaxy (i.e.\ the attenuation curve) depends not only on the extinction curve
but also on the spatial distribution of dust and stars
\citep[e.g.][]{2001PASP..113.1449C, 2005MNRAS.359..171I, 2018ApJ...869...70N}.
However, we would also like to predict the statistical properties of galaxies
at different redshifts, and for this, a full cosmological simulation
with a large box size
is more desired, in addition to the limited galaxy sample in zoom-in simulations. 

Recently, \citet{2019MNRAS.484.1852A} calculated the dust emission properties
by post-processing their cosmological simulation. Because of low spatial resolution,
they applied a one-zone calculation for the radiative properties. 
They succeeded in predicting the statistical properties of dust emission such as the
IR luminosity function, IRX--$\beta$ (IR-to-UV luminosity ratio vs.\
UV spectral slope) relation, etc. Indeed, for the IRX--$\beta$ relation, the
extinction curve (more precisely, attenuation curve; \citealt{2018MNRAS.474.1718N}) is
important. Since \citet{2019MNRAS.484.1852A} adopted
the two-size approximation, their extinction curves only had two degrees of freedom,
so that their predictive power of extinction curves was limited. By calculating the
full grain-size distribution in the future, 
we will obtain a better prediction for extinction curves also in 
cosmological simulations.

\section{Conclusion}\label{sec:conclusion}

We performed a hydrodynamic simulation of an isolated disc galaxy with
the evolution of full grain size distribution. We solved the time evolution of grain size distribution
caused by stellar dust production, SN destruction, shattering, accretion, and coagulation
on each gas (SPH) particle in the \textsc{gadget3-osaka} hydrodynamic simulation.
Each of the dust evolution processes above is treated in a consistent manner with the
local physical conditions (gas density, temperature, metallicity, etc.) of the gas particle.
As a consequence, we obtain the spatially resolved information on the grain size distribution
at each stage of galaxy evolution.

For the total dust abundance, our simulation reproduces the relation between
dust-to-gas ratio and metallicity, as well as the radial profiles of dust-to-gas ratio and dust-to-metal
ratio in nearby galaxies. 
We also confirm that the obtained results for the total dust
amount is consistent with our previous simulation (A17) which adopted the two-size
approximation. Therefore, we conclude that our models give reasonable results 
as far as the total dust abundance is concerned.

For the evolution of grain size distribution, we obtain the following results, 
taking advantage of the spatially resolved information. 
The grain size distribution is dominated by large grains in the earliest phase 
($t\lesssim 0.3$\,Gyr) in the entire disc.
In the intermediate stage ($t\sim 1$\,Gyr), 
small grains are more abundant in the dense, metal-enriched ISM,
where shattering and accretion can efficiently 
{ form a bump (second peak) at $a\sim 0.01\,\mu$m}.
Therefore, the extinction curves are steeper in the dense ISM than in the diffuse ISM. 
At a later stage ($t\gtrsim 3$\,Gyr), 
the relative abundance of small grains to large grains is higher 
in the diffuse ISM than in the dense ISM.
This is because higher dust abundances in the dense ISM are favourable for coagulation,
which creates large grains by depleting small grains. Accordingly,
the extinction curve becomes flatter in the dense ISM at later times.  
The grain size distribution approaches to the MRN distribution
at $t\gtrsim 3$\,Gyr, and the Milky Way extinction curve is reproduced well.

The grain size distribution depends also
on the position in the galaxy. At young ages
($t\lesssim 1$\,Gyr), the small-grain abundance is most enhanced in the 
central region (at small galactic radii $R$). This is because the highest dust abundance in the
central part leads to the most active shattering and accretion. 
In contrast,  the small-grain abundance is the lowest at small $R$ 
at a later stage ($t\gtrsim 3$\,Gyr) because of coagulation.
Accordingly, extinction curves are the steepest at small $R$ at young ($t\lesssim 1$\,Gyr)
ages, while they are the flattest at small $R$ at old ($t\gtrsim 3$\,Gyr) ages.
Since the above change in the grain size distribution occurs { at} $t\sim 1$ Gyr, the age and density dependence of grain size distribution has a significant impact on the extinction curves even at high redshift.

We emphasize that all the above results are consistent with the results obtained by
our simpler grain evolution model using the two-size approximation
\citep[e.g.][]{2018MNRAS.478.4905A,2019MNRAS.485.1727H,2019MNRAS.484.1852A}.
This not only indicates the robustness of the results in this paper but also supports our previous
results. We should, however, emphasize that the full grain size distribution is essential in
predicting observable quantities such as extinction curves and dust emission properties
(radiation transfer effects).

\section*{Acknowledgements}

{We thank the anonymous referee for useful comments.}
We are grateful to Ikkoh Shimizu for
useful discussions and comments regarding technical issues in the simulations.
We acknowledge Volker Springel for providing us with the original version 
of \textsc{gadget-3} code.
Numerical computations were carried out on Cray XC50 at the Center
for Computational Astrophysics (CfCA), National Astronomical Observatory of Japan,
Cray XC40 at the Yukawa Institute Computer Facility in Kyoto University,
and XL at the Theoretical Institute for
Advanced Research in Astrophysics (TIARA) in Academia Sinica.
HH thanks the Ministry of Science and Technology for support through grant
MOST 105-2112-M-001-027-MY3, MOST 107-2923-M-001-003-MY3 (RFBR 18-52-52-006),
and MOST 108-2112-M-001-007-MY3.
KN acknowledges the support from the JSPS KAKENHI Grant Number JP17H01111, 
as well as the travel support from the Kavli IPMU, World Premier Research Center Initiative (WPI), where part of this work was conducted.

\bibliographystyle{mnras}
\bibliography{ken}

\bsp	
\label{lastpage}
\end{document}